\documentclass[12pt]{article}

\usepackage{epsfig,amsfonts,amssymb}
\def \lsim{\mathrel{\vcenter
     {\hbox{$<$}\nointerlineskip\hbox{$\sim$}}}}
\def \gsim{\mathrel{\vcenter
     {\hbox{$>$}\nointerlineskip\hbox{$\sim$}}}}

\newcommand{\beq}{\begin{equation}}
\newcommand{\eeq}{\end{equation}}
\newcommand{\beqa}{\begin{eqnarray}}
\newcommand{\eeqa}{\end{eqnarray}}
\newcommand{\beqar}{\begin{eqnarray*}}
\newcommand{\eeqar}{\end{eqnarray*}}
\newcommand{\eg}{{\it e.g.,}\ }
\newcommand{\ie}{{\it i.e.,}\ }
\newcommand{\reef}[1]{(\ref{#1})}

\begin{document}

\thispagestyle{empty}

\hfill{}

\hfill{}

\hfill{}

\hfill{hep-th/0512274}

\vspace{32pt}

\begin{center}
\textbf{\Large Non-perturbative materialization of ghosts}\\

\vspace{40pt}

Roberto Emparan$^{a,b}$ and
Jaume Garriga$^a$

\vspace{12pt}
$^a$\textit{Departament de F{\'\i}sica Fonamental}\\
\textit{Universitat de
Barcelona, Diagonal 647, E-08028, Barcelona, Spain}\\
\vspace{6pt}
$^b$\textit{Instituci\'o Catalana de Recerca i Estudis Avan\c cats (ICREA)}\\
\vspace{6pt}
\texttt{emparan@ub.edu, garriga@ifae.es}
\end{center}

\vspace{40pt}

\begin{abstract}

In theories with a hidden ghost sector that couples to visible
matter through gravity only, empty space can decay into ghosts and
ordinary matter by graviton exchange. Perturbatively, such
processes can be very slow provided that the gravity sector
violates Lorentz invariance above some cut-off scale. Here, we
investigate non-perturbative decay processes involving ghosts,
such as the spontaneous creation of self-gravitating lumps of
ghost matter, as well as pairs of Bondi dipoles (\ie lumps of
ghost matter chasing after positive energy objects). We find the
corresponding instantons and calculate their Euclidean action. In
some cases, the instantons induce topology change or have negative
Euclidean action. To shed some light on the meaning of such
peculiarities, we also consider the nucleation of concentrical
domain walls of ordinary and ghost matter, where the Euclidean
calculation can be compared with the canonical (Lorentzian)
description of tunneling. We conclude that non-perturbative ghost
nucleation processes can be safely suppressed in phenomenological
scenarios.

\end{abstract}

\setcounter{footnote}{0}

\newpage
\section{Introduction}

Recently it has been suggested that the smallness of the observed
cosmological constant can be attributed to an approximate ``energy
symmetry" \cite{kasu}. The idea is that Nature is endowed with an
exact copy of the matter sector, but with an overall minus sign in
the action \cite{li,kasu},
\begin{equation}
{\cal L} = \sqrt{-g}\{ M_P^2 R + {\cal L}_{matt}(\psi,g) - {\cal
L}_{matt}(\hat\psi, g)+...\}. \label{esym}
\end{equation}
Here, $g$ is the metric, $\psi$ are ordinary matter fields
(including those of the Standard Model), and $\hat \psi$ are the
ghost fields. Energy parity is defined by $\psi \to \hat \psi,
\quad \hat\psi\to \psi, \quad g\to g.$ Ignoring gravity, the
Hamiltonian $H$ transforms as
\beq H\to -H\,. \label{htomh}
\eeq
The
vacuum state $|0\rangle$ is defined as parity invariant, and from
(\ref{htomh}) the corresponding vacuum energy vanishes to all
orders in perturbation theory. However, gravity breaks the energy
symmetry and a cosmological constant is induced. It was argued in
\cite{kasu} that the magnitude of this vacuum energy can be
comparable to the one suggested by observations provided that the
gravitational cut-off scale $\mu$ is sufficiently low (lower than
the inverse of 30 microns, about an order of magnitude beyond the
reaches of ongoing short distance probes of gravity).

``Phantom" matter has also been invoked in phenomenological
studies of dark energy \cite{de,ca,cline}, as a way of obtaining an
effective equation of state parameter $w < - 1$. This violates all
the standard energy conditions, but is not at all disfavoured by
observations. A ghost sector is also present in the recently
proposed B-inflation, based on effective theories with only second
derivatives of a scalar field \cite{an}.

In all of these cases, the Hamiltonian is unbounded below, and
disaster would follow unless one postulates that the Lorentz
symmetry is broken at a certain energy scale \cite{cline}. The reason is
simple. In the model (\ref{esym}), empty space can decay into a
pair of $\psi$ ordinary particles and a pair of $\hat \psi$ ghost
particles
\begin{equation}\label{decay1}
|0 \rangle \to \psi\psi\hat\psi\hat\psi \label{1}
\end{equation}
(if particles are charged, one in each pair should be understood as the
antiparticle). Let the momenta of the ordinary particles be $p_1$ and
$p_2$, and the momenta of the ghost particles $k_1$ and $k_2$. From
translation invariance, the decay amplitude takes the form
$\langle p_1,p_2,k_1,k_2|0\rangle={\cal A}(p_1,p_2,k_1,k_2)\
\delta^{(4)}(p_1+p_2+k_1+k_2)$, which after integration over
external momenta leads to the vacuum decay rate per unit volume
\begin{equation}
\Gamma = \int d^4 P\ \gamma(P), \label{3}
\end{equation}
where $\gamma(P)= \int d {\tilde p_1}\ d{\tilde p_2}\ d {\tilde
k_1}\ d{\tilde k_2}\ |{\cal A}|^2 \ \delta^{(4)}(P+k_1+k_2)\
\delta^{(4)}(P-p_1-p_2)$. In a Lorentz invariant theory,
$\gamma(P)$ is just a function of $s=-P_\mu P^\mu$, and Defining
$\vec v = s^{-1/2} \vec P$, we have
\beq
\Gamma = \int ds\ s\
\gamma(s) \int {d^3 \vec v\over 2 \sqrt{1+\vec v^2}}.
\eeq
Physically, the last integral corresponds to the fact that there
is no preferred reference frame, and the total momentum $P^{\mu}$
of the pair of particles (or the pair of ghosts) is equally likely
to fall anywhere on the mass-shell of radius $s^{1/2}$. Particles
only interact with ghosts gravitationally, and so the momentum
$P^{\mu}$ is transferred by gravitons. The decay rate is in
principle infinite (due to the mass-shell integral) but it can be
rendered finite if we postulate that Lorentz invariance is broken
in the gravitational sector at some scale ${\cal E}$ \cite{cline}. The
remaining integral over $s$ can be finite in a theory where
gravity becomes soft at a certain cut-off scale $\mu$, as it is in
fact assumed. The process becomes completely negligible if ${\cal
E}$ is comparable to the cut-off scale $\mu\lsim(30\,
\mu\mathrm{m})^{-1}$ discussed above \cite{kasu}. Similarly, empty
space can decay to ghosts $\hat\psi$ and gravitons
$h$,\footnote{Ref.~\cite{kasu} actually considered the decay $|0
\rangle \to h^*\hat\psi\hat\psi$, where $h^*$ is an ``excited"
(soft-scale) graviton, which is a more dominant process than
\reef{decay2}. We shall not consider the non-perturbative analogue
of this process, since it cannot be described in terms of the low
energy effective action (\ref{esym}).}
\begin{equation}\label{decay2}
|0 \rangle \to hh\hat\psi\hat\psi. \label{2}
\end{equation}
In this case, the integrals over the momenta of the external
gravitons must be cut-off at the Lorentz violating energy scale
${\cal E}$. In this way, the vacuum can be made sufficiently
stable to perturbative decay processes, in spite of the ghosts
\cite{kasu,ca}.

Although perturbative processes may be suppressed by the
Lorentz-violating physics, it is conceivable that non-perturbative
processes may quickly destabilize the present vacuum, through the
production of lumps of non-relativistic ghost matter. The purpose
of the present paper is to investigate the non-perturbative analogues of
(\ref{1}) and (\ref{2}). Decays that proceed via non-perturbative
tunneling are typically slower than their perturbative counterparts,
but when ghosts are involved there are several reasons why this is
not so obvious.

In accordance with the equivalence principle, a lump of ghost matter
tends to fall towards the potential well created by a positive energy
object. On the other hand, the repulsive gravitational field it produces
tends to push the positive energy object away. It has been known for
some time that this leads to a runaway behaviour, where the positive
energy object is chased after by the ghost
\cite{bondi,iskh,bonnor,gibbons}, with a constant acceleration. Such
configuration is known as a {\em Bondi dipole} \cite{bondi}. As we shall
see, such self-accelerating solutions can be continued to the Euclidean
section, leading to a semiclassical description of the spontaneous
nucleation of pairs of Bondi dipoles. This would be the non-perturbative
analogue of (\ref{1}). A simple estimate (which we will confirm by
rigorous calculation) gives the Euclidean action of this process
as $\sim m_+ d$, where $d$ is the size of the dipole and $m_+$ is the
mass of its ordinary positive-mass component. Even if we impose
$d\gsim\mu^{-1}$, we see that the action can dangerously approach a value of
order one if the Compton wavelength of the particles produced is also
close to the gravitational cutoff.

The analogue of (\ref{2}) is the pair creation of self-gravitating lumps
of ghost matter, which repel each other. The possibility of this process
is suggested by the following weak field argument. The interaction
energy of two ghost particles at rest, with identical mass $m<0$, is
given by $E_{grav}=G m^2/ 2 r$. Here $G$ is Newton's constant and $2 r$
is the distance between the masses. For $r= -G m/4$ the positive
gravitational energy is equal to minus the rest mass energy of the pair
$E_{grav}= -2 m$, so this configuration can in principle pop out of the
vacuum without violating energy conservation. Also, the initial
acceleration of each particle is given by $a= 1/r$, suggesting that
there is a Euclidean solution where the ghost matter runs around a
circle of radius $1/a$. Note, however, that for $r \sim -G m$ the
gravitational field is of order one, and non-linear gravity must be
taken into account. The corresponding instantons still exist, and have
interesting peculiarities which make their interpretation non-trivial.
First of all, they can produce a topology change, and second, the
corresponding Euclidean action (defined as the bounce action minus the
background action) can be {\it negative}.

To shed some light into the meaning of such peculiarities, we
shall first consider the simpler example of vacuum decay in a
theory where the matter and the ghost sectors support domain wall
solutions. In this case, the process of spontaneous nucleation of
concentrical spherical domain walls of ordinary and ghost matter
chasing after each other is the analogue of (\ref{1}). The
analogue of (\ref{2}) is the spontaneous creation of spherical
domain walls of ghost matter. If the cosmological constant is
exactly vanishing, the instanton for the latter process changes
the topology of space and a new boundary appears inside of the
domain wall. This makes the calculation of the corresponding
Euclidean action somewhat ill-defined. On the other hand, in the
presence of a small cosmological constant (such as the one present
in our universe), the topology does not change at all and the
Euclidean action can be calculated unambiguously. Interestingly,
it turns out to be negative. These features are quite similar to
what happens in the case of pair creation of lumps of ghost
matter, but the advantage here is that the geometry is much
simpler, and the Euclidean calculation can be compared with a
canonical (Lorentzian) description of tunneling.

\bigskip

\noindent The plan of the paper is the following. Since the subject of
tunneling in theories with ghosts is fraught with many subtleties, we
have developed it in quite some detail in Sections \ref{sec:ghostwalls}
through \ref{sec:ghostpairs}. The specific application of our results to
the energy-symmetric scenario of \cite{kasu} is then discussed in the
concluding Section \ref{sec:disc}, to which phenomenologically-minded
readers might want to jump directly if not interested in the theory of
ghost tunneling.

The more technical sections \ref{sec:ghostwalls}--\ref{sec:ghostpairs}
consist of a discussion of: the spontaneous nucleation of domain walls
(Section \ref{sec:ghostwalls}); the pair creation of Bondi dipoles, \ie
the non-perturbative analogue of \reef{decay1} (Section
\ref{sec:bondidip}, where also we briefly review several technical
aspects of the axisymmetric class of solutions for the benefit of
readers unfamiliar with them); the spontaneous creation of
self-gravitating lumps of ghost matter, \ie the non-perturbative
analogue of \reef{decay2} (Section \ref{sec:ghostpairs}). Some details of
the canonical WKB construction of tunneling paths are deferred to an Appendix.

\section{Ghosts through the tunnel}

\label{sec:ghostwalls}
In field theory, semiclassical tunneling rates are usually estimated
through the expression
\begin{equation}
\Gamma \sim e^{-I_E}. \label{rate}
\end{equation}
Here, $I_E$ is the action of the Euclidean instanton describing
the decay, minus the action of the background, and we have omitted
the prefactor arising from integration of fluctuations around
these solutions. For ordinary fields, with the rotation $t=i t_E$,
the Euclidean action is positive definite, $I_E>0$, and the above
formula gives an exponential
suppression in the limit when the semiclassical approximation is valid,
$I_E\gg1$. With the same rotation $t=i t_E$, the Euclidean action for
ghost matter is negative definite, $I_E<0$, and from a naive application
of (\ref{rate}) one may be tempted to conclude that ghosts lead to
catastrophic decay rates. However, this conclusion would be premature.
Rather, the problem is that the Euclidean path integral is ill-defined:
in order to make it convergent, ordinary matter and ghosts would require
opposite Wick rotations. Hence, the standard Euclidean methods are not
directly applicable in the present context.

For instance, in the limit when gravity is neglected $G\to 0$, the
theory (\ref{esym}) is symmetric under energy parity. In this limit,
ordinary and ghost matter are decoupled and have exactly the same
dynamics. Note that, as a consequence, in \eg Schwinger pair
production, it is just as hard to screen a ghost electric field by
nucleation of charged ghosts, as it is to screen an ordinary electric
field by nucleation of ordinary charged particles. So in both cases,
instanton processes must be exponentially suppressed, as
\begin{equation}
\Gamma \sim e^{-|I_E|}. \label{rate2}
\end{equation}
Thus, when gravity is switched off, nonperturbative processes in
the ghost sector will not bring any disaster, even if with the
standard rotation the Euclidean action is negative.

When gravity is switched on, both sectors are coupled and, as
mentioned above, the standard Euclidean methods do not apply.
Hence, we should try to develop some understanding of the problem
from the canonical approach to tunneling. In the WKB
approximation, the wave function is of the form $\Psi \sim A
\exp(iW) + B \exp (-iW)$, where for a simple quantum mechanical
system $W(q)= \int^q p(q')dq'$, and the integral is taken along a
semiclassical trajectory. Under the barrier, the momentum $p$ is
imaginary and $\Psi$ becomes a superposition of growing and
decaying exponentials. We are interested in the situation where
the wave function is outgoing after tunneling, so generically we
have a comparable contribution of the growing and decaying modes
at the turning point after the barrier. This means that the
amplitude $\Psi_a$ {\em after} the barrier is exponentially
smaller than the amplitude $\Psi_b$ {\em before} the barrier
\beq
|\Psi_a/\Psi_b|^2 \sim \exp(-2|\Delta W|)
\eeq
Here, $\Delta W$ is
the difference in $W$ evaluated at the two turning points. From
this perspective, we should expect that a tunneling process is
suppressed, whether it involves ghosts or not.

In order to gain some intuition on this problem, we shall consider
in the following subsections the spontaneous nucleation of
spherical domain walls. The dynamics of domain walls is
sufficiently simple to be discussed in the canonical formalism.
First, in Subsections 2.2, 2.3 and 2.4 we shall describe some
instanton solutions which in the standard interpretation would
correspond to the nucleation of walls in flat and in deSitter
space. In Subsection 2.4 we consider the same processes in the
canonical approach, without reference to Euclidean methods.
Finally, in Subsection 2.5, and in the light of the examples
considered, we ellaborate on the possible relation between the
Euclidean action and the nucleation rates.

\subsection{Nucleation of diwalls from flat space}

The gravitational field of an ordinary domain wall is repulsive
\cite{av}. On the other hand, a ghost wall will
be attractive, and we may expect to find solutions where a wall
of ghost matter is chased after by a wall of usual matter. By
analogy with the Bondi dipoles discussed in the introduction, we
may call such a configuration a {\em diwall}.

\begin{figure}%[ht]
%%\vspace{0.5in}
\begin{center}\leavevmode  %
\epsfxsize=14cm \epsfbox{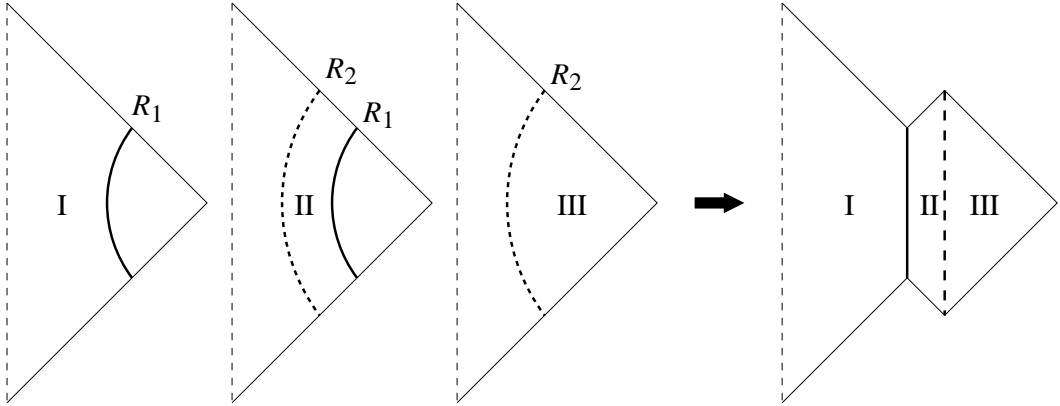}
\end{center}
%%\vspace{-0.5in}
\caption{\small Conformal diagrams for the construction of the {\em
diwall} solution. Three different regions (I, II, III) of Minkowski
space are cut out at given radii $R_1>R_2$, and pasted to form a wall of
positive tension (thick solid line) at $R_1$ and a ghost wall of negative
tension (thick dashed line) at $R_2$. In the final diagram, moving
towards the right corresponds to increasing radius in I and III, and to
decreasing radius in II.
}
\label{fig:diwall}
\end{figure}

Domain walls are rather easy to treat as distributional sources in
General Relativity. Their effect is a discontinuity in the
extrinsic curvature $K_{ab}$ accross the worldsheet \cite{di}
\beq [K_{ab}]= -4\pi G \sigma \gamma_{ab},\label{xtr}\eeq where
$\sigma$ is the tension of the wall and $\gamma_{ab}$ is the
induced metric. Consider a Euclidean spherically symmetric
solution with a wall of positive tension $\sigma_1$ and a wall of
negative tension $-\sigma_2$. The metric takes the form
\begin{equation}
ds^2= dy^2 + R^2(y) d\Omega_3, \label{metwall} \end{equation} where
$d\Omega_3$ is the line element on the 3-sphere and $y$ is a radial
coordinate. Outside the sources, the metric is flat, and the warp
factor $R(y)$ is piecewise linear with slope $dR/dy=\pm 1$. At the
location of the sources, the slope is discontinuous, to account
for the jump in the extrinsic curvature. Hence, starting from the
center of symmetry at $y=0$ the solution is given by (see
fig.~\ref{fig:diwall})
\begin{eqnarray}
R(y) &=& y   \quad\quad(0 < y < R_1), \nonumber\\
R(y) &=& 2 R_1 - y\quad\quad(R_1 < y < 2 R_1 - R_2),\nonumber\\
R(y) &=& 2 R_2 - 2 R_1 +y \quad\quad(2 R_1 -R_2 < y < \infty).
\label{pm}
\end{eqnarray}
Eq.~(\ref{xtr}) demands that $R_1 = 1/2\pi G \sigma_1$ and $R_2 = 1/2\pi
G \sigma_2$. The radius $R(y)$ of the 3-spheres
increases up to $R_1$, backtracks to $R_2$, and then increases to
infinity. Note that $R_2<R_1$, which requires $\sigma_2 >
\sigma_1.$ For our illustrative purposes, we shall simply assume
that the theory supports domain walls satisfying this inequality.

The solution Eq.~(\ref{pm}) is perfectly regular and
asymptotically flat. It can be thought of as a semiclassical
trajectory which interpolates between flat empty space ${\mathbb R}^3$ (at
infinity), and the equatorial slice of the metric (\ref{metwall}).
This ``turning point" slice contains two concentrical domain walls
of radii $R_1$ and $R_2$.

After nucleation, the evolution is given by the analytic
continuation of (\ref{metwall}) to Lorentzian time, which converts
the 3-spheres into time-like hyperboloids. In particular, the
positive and negative tension walls expand with constant proper
acceleration $1/R_1$ and $1/R_2$ respectively. Due to the peculiar
``backtracking" form of the metric between $R_1$ and $R_2$, it is
the wall of larger radius (and positive tension) which chases
after the one of smaller radius (and negative tension) as they
both expand. This is as it should be, since the wall of positive
tension is repulsive and the other one is attractive.

The Euclidean action can be easily calculated from
\beq
I_E =
-{1\over 16\pi G}\int d^4 x \sqrt{g}\; {\cal R} + \sum_i s_i
\sigma_i \int d^3\xi \sqrt{\gamma}, \label{euact}
\eeq
where $s_i$
is the sign of the wall tension (and $\sigma_i$ is as usual its
absolute value). On shell, the Ricci scalar is related to the
source \cite{bgv}
\beq
\sqrt{g}\; {\cal R}=24 \pi G\sigma \int
d^3\xi \sqrt{\gamma}\;\delta^4(x-x(\xi)),\label{aux}
\eeq
and
integrating over the volume of the 3-spheres, we easily find
\beq
I^{diwall}_E = {1\over 8\pi G^3} (\sigma_2^{-2}-\sigma_1^{-2}) <
0.\label{diwact}
\eeq
Note that the Euclidean action is negative.
In the limit $\delta\sigma=\sigma_2-\sigma_1 \ll \sigma$, we have
$|I_E| \sim (M_P^6/\sigma^2)(\delta\sigma/\sigma)$. This is likely
to be very large, unless the wall tensions are very nearly
degenerate, or unless they are very close to the Planck scale
$M_p$.

The diwall instanton (\ref{pm}) is an analogue of the process
(\ref{1}), where ghost and ordinary matter are created from the
vacuum. Let us now investigate the analogue of (\ref{2}).

\subsection{Ghost walls from flat space?}

\begin{figure}%[ht]
%%\vspace{0.5in}
\begin{center}\leavevmode  %
\epsfxsize=10cm \epsfbox{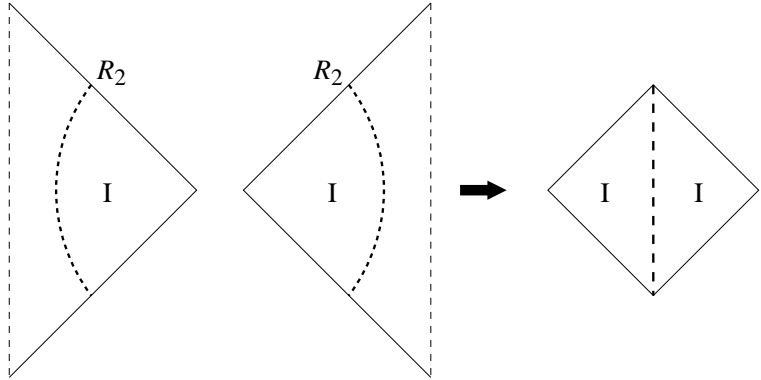}
\end{center}
%%\vspace{-0.5in}
\caption{\small Construction of the ghost wall spacetime. Observe the
presence of two disconnected asymptotic boundaries.
}
\label{fig:ghostwall}
\end{figure}

An instanton for a single spherical domain wall of negative tension in an
asymptotically flat space can be constructed along the
same lines as in the previous subsection. The warp factor is now
given by (see fig.~\ref{fig:ghostwall})
\begin{equation}
R(y) = |y|+R_2   \quad\quad(-\infty< y < \infty). \label{m}
\end{equation}
From large radii ($y\to \infty$), the worldsheet is seen as a
spherical object of radius $R_2$ embedded in an otherwise flat
Euclidean space. However, if we cross the worldsheet towards
negative $y$, we discover that the radius does not shrink to zero.
Rather, it grows as we go in. This ``throat" is of course an
identical copy of the geometry for $y>0$. The topology of the
instanton is not ${\mathbb R}^4$, but ${\mathbb R} \times S^3$, and the solution has
two disconnected boundaries, one at $y\to \infty$ and another one
``inside" the domain wall, at $y\to -\infty$.

Calculating the Euclidean action as we did in the previous
subsection would yield
\beq
I_E = \frac{1}{\pi G^3 \sigma_2^2} >
0. \label{as}
\eeq
But the fact that the instanton has different
topology than the background should give us pause. The
Gibbons-Hawking boundary term $I_E^b= -(8\pi G)\int
d^3\xi\sqrt{\gamma}\; K$ at the new inner boundary ($y\to
-\infty$) is given by $I_E^b =-3\pi R_c^2/4G < 0$, where $R_c$ is
some large cut-off radius. It is unclear what subtraction (if any)
should be performed on this divergent contribution, since the
original flat space background does not include such inner
boundary. Also, from a canonical point of view, it doesn't seem
possible to go from the original ${\mathbb R}^3$ to the final turning point
geometry (with two asymptotically flat regions) by continuous
slicing of the instanton. Because of that, the interpretation of
the solution (\ref{m}) as a semiclassical path describing the
nucleation of a ghost wall remains unclear.

Fortunately, the situation is clarified by considering nucleation
from a vacuum with a small cosmological constant.

\subsection{Walls from deSitter space}
\label{sec:dswall}

As mentioned in the introduction, gravity breaks the energy
symmetry (\ref{htomh}). This may give rise the small observed
cosmological constant $\Lambda>0$. In this case, the Euclidean
background is the 4-sphere of radius $H^{-1}=(3/\Lambda)^{1/2}$,
with warp factor
\beq
R_b(y)=H^{-1} \sin(H y), \label{s4} \quad\quad
(0\leq Hy \leq \pi).
\eeq
Nucleation of positive tension domain
walls in deSitter has been thoroughly studied in Ref.~\cite{bgv}.
The relevant instanton is constructed from two caps of the
4-sphere, joined at the worldsheet of the domain wall (which is an
$S^3$),
\begin{eqnarray}
R_a(y)&=&H^{-1} \sin(H y), \quad\quad
(0\leq
Hy\leq \chi_0),\nonumber\\
R_a(y)&=&H^{-1} \sin(2\chi_0 - H y), \quad\quad (\chi_0\leq Hy\leq
2\chi_0).\label{caps}
\end{eqnarray}
The angular span $\chi_0$ of
the spherical cap is determined through the junction condition
(\ref{xtr}) \beq \tan\chi_0 = {H\over 2\pi G\sigma_1}, \label{ang}
\eeq where $\sigma_1$ is the tension of the domain wall. Thus, the
radius of the worldsheet is given by
\begin{equation}
R_w= R_a(y_0) = H^{-1} \sin\chi_0 = {1\over \sqrt{H^2 + (2\pi
G\sigma_1)^2}}. \label{rw}
\end{equation}
The Euclidean action can be calculated from (\ref{euact}) with an
additional vacuum energy term $\frac{1}{8\pi G}\int d^4 x\; \sqrt{g}
\;\Lambda$. On shell, we can use (\ref{aux}) as well as the bulk
equations of motion ${\cal R}= 4\Lambda$.

Properly speaking, this instanton would describe the creation from
nothing of a deSitter space containing a wall. If we want to
describe instead the transition from empty deSitter to deSitter
with a wall, then we must face the well-known problem that no
non-singular instanton mediates between both spacetimes. For the
time being, we will follow the common procedure of calculating the
action for the whole process by subtracting the action $I_{dS}$ of
the background deSitter,
\beq\label{dsaction}
I_{dS}=-\frac{\pi}{GH^2}\,.
\eeq
This will be justified in more detail in the next subsection.
Some straightforward algebra
leads to \cite{bgv}
\beq
I^+_E= {2\pi^2\sigma_1 \over H^2
\sqrt{H^2+(2\pi G\sigma_1)^2}}>0.\label{actp}
\eeq

Let us now consider the nucleation of negative tension walls in de
Sitter. In fact, Eqs. (\ref{ang}-\ref{actp}) still hold, with the
replacement $\sigma_1 \to -\sigma_2$. The angle $\chi_0$, determined by
\beq
\tan\chi_0 = -{H\over 2\pi G\sigma_2},
\eeq
will be larger than $\pi/2$, so instead of having
two spherical caps glued to the worldsheet,
it will now be two capped spheres which are
glued. The corresponding euclidean action is negative
\beq
I_E^- = -{2\pi^2\sigma_2 \over H^2 \sqrt{H^2+(2\pi G\sigma_2)^2}}<
0.\label{actn}
\eeq
Unlike the case of ghost walls from flat space
discussed in the previous
subsection, the topology of the instanton is here the same as for the
deSitter background, and no additional boundaries appear.
Note that in the limit of a small cosmological constant $H\to
0$, the action (\ref{actn}) tends to negative infinity, unlike the
naive result (\ref{as}).

In the limit of small tension, where the
walls do not deform the geometry of deSitter,
the equations of motion for positive and negative tension are exactly
the same, and therefore it is expected on
general grounds that it is just as hard to create a positive
tension wall as it is to create a negative tension one. Here, we find
that $|I_E|$ has the same form in both cases, which supports the
use of Eq.~(\ref{rate2}) [rather than Eq.~(\ref{rate})] for the
calculation of the nucleation rates.

It is interesting to observe also that
$|I_E|$ has the same form for both signs of the tension
even when this tension is large and the background geometry is
considerably deformed (note that the shape of the instantons for
positive and negative tension walls is quite different in this case).

\subsection{Canonical approach}

In this subsection, we shall consider the processes discussed in
the previous ones in the canonical WKB approach, without reference
to Euclidean methods.

\begin{figure}%[ht]
%%\vspace{0.5in}
\begin{center}\leavevmode  %
\epsfxsize=10cm \epsfbox{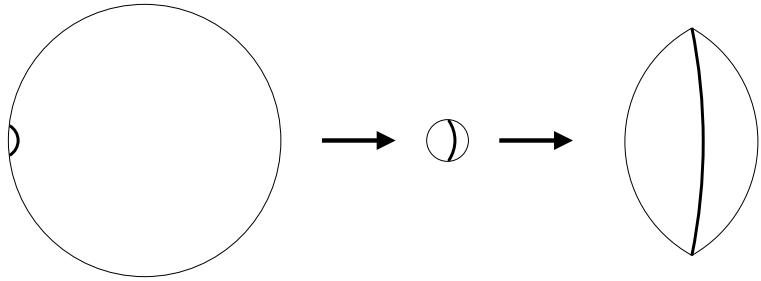}
\end{center}
%%\vspace{-0.5in}
\caption{\small WKB path for the spontaneous nucleation of a domain wall
of positive tension in deSitter. The leftmost figure represents the
``initial" classical turning point 3-geometry before tunneling. It
consists of a small domain wall of infinitessimal mass $M$. Inside the
wall, the metric is a small cap of a 3-sphere of radius $H^{-1}$ (which
looks almost flat on the scale of the radius of the wall). Outside the
wall, the geometry is a $t=\mathrm{constant}$ section of a Schwarzschild-deSitter
metric with a very small mass. Classically, the small wall would shrink
under its own tension and form a small black hole. However, it has a
certain probability for tunneling into a big wall, of size comparable to
the cosmological horizon, through a sequence of interpolating
``underbarrier" 3-geometries (whose construction is described in
Subsection 2.4 and in the Appendix). The ``final" turning point geometry
is represented in the rightmost figure. In the limit when $M\to 0$, this
final turning point geometry consists of two large caps of a 3-sphere of
radius $H^{-1}$ glued at the wall. Note that the intermediate
interpolating 3-geometries in such path shrink to a very small size,
just a few times larger than the size of the initial domain wall. In the
limit $M\to 0$ this intermediate geometry would shrink to nothing. The
initial and final geometries can also be thought of as equatorial slices
of the deSitter instanton and of the domain wall instanton [given in
Eq.~(\ref{caps})] respectively. }
\label{fig:wallpath}
\end{figure}

\begin{figure}%[ht]
%%\vspace{0.5in}
\begin{center}\leavevmode  %
\epsfxsize=14cm \epsfbox{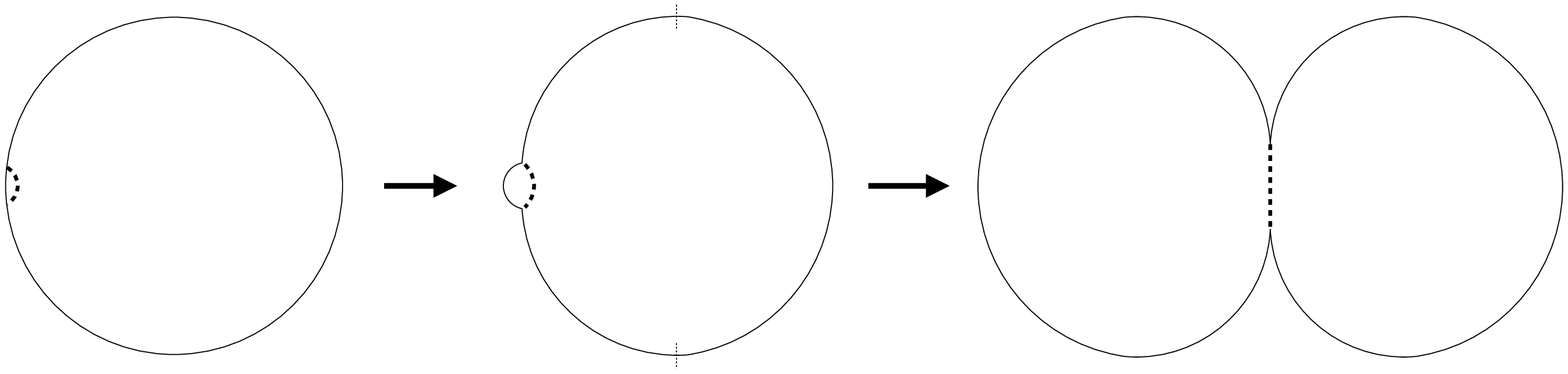}
\end{center}
%%\vspace{-0.5in}
\caption{\small Same as Fig.~\ref{fig:wallpath} for a ghost domain wall.
The final turning point geometry consists now of two capped 3-spheres of
radius $H^{-1}$, which correspond to the equatorial slice of the
instanton (\ref{caps}). As shown in the Appendix, another difference
with the case of positive tension walls is that the volume of the
intermediate three-geometry in the semiclassical path never shrinks to a
very small size, even in the limit $M\to 0$.
}
\label{fig:ghostpath}
\end{figure}

 Let us consider the system of a
spherically symmetric domain wall in the presence of a
cosmological constant $\Lambda \equiv 3H^2$. Inside the wall, the metric
is just empty deSitter space, whereas outside the wall, and by
Birkhoff's theorem, the metric is Schwarzschid-deSitter, characterized
by a mass parameter $M$. This system has a single
degree of freedom, which is the radius of the wall. A domain wall
of very small radius, and correspondingly very small mass $M$, can
tunnel to a big domain wall of size comparable to the cosmological
horizon (while the mass parameter $M$ remains of course a small
constant). In the limit when $M \to 0$, this process corresponds
to the spontaneous nucleation of a large domain wall (from an
infinitessimally small seed) in an otherwise empty deSitter
space. This is schematically represented in Figs.~\ref{fig:wallpath} and
\ref{fig:ghostpath} for the case of ordinary walls and ghost walls
respectively.

In order to describe this process, we shall closely follow the procedure
developed in Ref.~\cite{fmp}.
The action is given by
\beq
I={1\over 16\pi G} \int d^4 x \sqrt{-g} {\cal R} - {\Lambda \over 8\pi
G} \int d^4 x \sqrt{-g}
 - {\sigma} \int d^3 \xi \sqrt{-\gamma}. \label{cact}
\eeq The spherically symmetric metric takes the form \beq ds^2 = -
(N^t)^2 dt^2 + L^2 (dr + N^r dt)^2+ R^2 d\Omega_2, \eeq where $N^t, L$
and $R$ are functions of $r$ and $t$, and $d\Omega_2$ is the metric on
the two-sphere. The action can be written as \beq I=\int dt\; p\;
\dot q + \int dr\ dt\ (\Pi_L \dot L + \Pi_R \dot R - N^t {\cal
H}_t - N^r {\cal H}_r), \eeq where $q(t)$ is the radial coordinate
$r$ of the wall at time $t$, the momenta
\begin{eqnarray}
p &=& -4\pi \sigma R^2 L^2 (\dot q + N^r)[N^{t\ 2}-L^2 (\dot q +
N^r)^2]^{-1/2}, \nonumber \\
\Pi_R &=& [(N^rLR)'-(LR)\dot{}\ ]/G N^t, \nonumber \\
\Pi_L &=& R\ (N^r R' - \dot R)/G N^t.
\end{eqnarray}
are conjugate to $q,L$ and $R$ respectively, and the Hamiltonian
densities are given in terms of the canonical variables as
\begin{eqnarray}
{\cal H}_t(r)&=&{GL\Pi_L^2\over 2 R^2}-{G\Pi_L\Pi_R\over R}+{1\over
2G}\left\{\left({2RR'\over L}\right)'-{R'^2\over L} - L\right\} +
{\Lambda LR^2\over 2G} + {E\over L}\ \delta(q-r), \\& & \nonumber \\
{\cal H}_r(r)&=& R'\ \Pi_R - L\ \Pi'_L -p\ \delta(q-r).
\label{piru}
\end{eqnarray}
Here $E={\rm sign}(\sigma)(p^2+16\pi^2\sigma^2L^2R^4)^{1/2}$, and
the primes indicate derivatives with respect to $r$. Time
derivatives of lapse and shift do not appear in the action, which
leads to the constraints $\Pi_{N^t}=\Pi_{N_r}=0$. When imposed on
the wave function, this leads to
$$
\Psi=\Psi[L,R,q].
$$
Derivative of the action with respect to lapse and shift gives the
constraints ${\cal H}_{t,r}=0$.

With the ansatz
$$
\Psi=e^{i W(L,R,q)},
$$
the WKB approximation is obtained from a solution of the Hamilton-Jacobi equations
$$
{\cal H}_{t,r}\left({\partial W\over \partial L},{\partial W\over
\partial R},{\partial W\over \partial q}, L,R,q\right)=0.
$$
We are interested in a solution which interpolates between suitable turning points.

Away from the source at $r=q(t)$, we may define
\beq M(r)= {G
\Pi_L^2\over 2R} +{R\over 2G}\left\{1-\left({R'\over
L}\right)^2-{\Lambda R^2\over 3}\right\}
\label{mass},
\eeq
and we
have $M'=- R' {\cal H}_t/L - {G\Pi_L {\cal H}_r/RL}=0$, where we
have used the Hamiltonian constraints. This implies that $M$ is
constant for $r\neq q$. At the turning point geometries the
momenta vanish, $\Pi_L=0$, and one has \beq \left({dR \over
dy}\right)^2 =  1-{2GM\over R} -{\Lambda R^2\over 3}, \eeq where
$dy=L dr$. This is of course the relation between the intervals of
proper length $dy$ and proper radius of the 2-spheres $dR$ in the
Schwarzschild-(A)dS geometry (in static coordinates). We are
mostly interested here in the case with $M=0$, and with
$\Lambda>0$, where turning point geometries $R_b(y)$ and $R_a(y)$
before and after the tunneling will be the equatorial slices of
the 4-sphere and of the domain-wall deSitter solution. The
functions $R_b(y)$ and $R_a(y)$ are given by Eqs. (\ref{s4}) and
(\ref{caps}) respectively.

{}From (\ref{mass}), we have \beq \Pi_L^2 = -{R^2\over
G^2}\left\{1+V(R)-\left({R'\over L}\right)^2\right\},\label{pil}
\eeq where we have introduced
$$
V(R)=-2GM/R -(\Lambda/3) R^2.
$$
At the wall $(r=q)$, the constraints are consistent with $L$ and
$R$ being continuous, while they imply the following
discontinuities for $\Pi_L$ and $R'$: \beq [\Pi_L] = -p/L,
\quad\quad [R'] = - G E/R. \label{pandq} \eeq [note that at the
turning point, the first equation is trivial since the momenta
vanish, and the second reproduces Eq.~(\ref{xtr})]. Moreover, from
Eq.~(\ref{piru}) we have
\beq
\Pi_R=L {\Pi_L'\over R'} + {p\over
R'} \delta(q-r), \label{pir}
\eeq
which by virtue of (\ref{pandq})
does not have delta-function contributions on the wall. Note that
from (\ref{pil}) and (\ref{pir}) we can write $\Pi_L$ and $\Pi_R$
in terms of $L(r)$ and $R(r)$. In particular
\beq
\Pi_R=
\left({L\over R}+{V'/2-(R'/L)'\over 1+V-(R'/L)^2}\right)\ \Pi_L,
\eeq
where $V'\equiv dV(R)/dR$ and $\Pi_L$ is given by
(\ref{pil}). Also, the momentum $p$ is determined by (\ref{pandq})
in terms of $L(r)$ and $R(r)$ in the neighborhood of $r=q$.

The solution of the Hamilton-Jacobi equation $W(L,R,q)$ satisfies
\beq
\delta W= p\ \delta q + \int (\Pi_L\ \delta L + \Pi_R\ \delta
R)\ dr,\label{HJ}
\eeq
for arbitrary variations $\delta q, \delta
L(r)$ and $\delta R(r)$. Under the barrier, where $R'^2 \leq L^2
(1+V)$, we may define \cite{fmp}
\beq
\Omega(L,R,q)=G^{-1} \int
dr\ \left( R L \sqrt{1+V-(R'/L)^2} - R\ R' \alpha
\right),\label{omega}
\eeq
where we have introduced the function
\beq
\alpha=\arccos\left(R'\over L\sqrt{1+V}\right).
\label{alpha}
\eeq
Here, the inverse cosine is defined in the range
$0\leq \alpha\leq \pi$ (so that the sine is positive). It is
straightforward to check that under arbitrary variations of the
arguments of $\Omega$, we have
\beq
\delta \Omega = \pm i \delta W
- G^{-1}\int dr\ {d\over dr} \left(\alpha R \ \delta R \right) +
G^{-1}[R' \alpha]R\ \delta q .\label{trt}
\eeq
The second term in
the right hand side arises from partial integration in the
variation with respect to $R$, and it is non-vanishing because
$R'$ is discontinuous across the wall. The square brackets in the
last term indicate the discontinuity (that is, the quantity
evaluated at $r=q+\epsilon$ minus the quantity evaluated at
$r=q-\epsilon$). This term arises from variation with respect to
$q$ of the second term in (\ref{omega}) (variation of the first
term in (\ref{omega}) produces the term $\pm i p \delta q$, which
has been included in $\pm i \delta W$). The last two terms in the
right hand side of (\ref{trt}) only depend on quantities evaluated
near the wall. Using $d\hat R = \delta R + R' \delta q$, where
$\hat R = R(r=q)$ is the proper radius of the wall, they can be
combined into $G^{-1} [\alpha] \hat R d\hat R$. Hence, we have
\begin{equation}
\pm\ i W(L,R,q) = \Omega - G^{-1}\int [\alpha]\hat R d\hat R.
\label{wsol}
\end{equation}
Since $\alpha$ is a function which depends on $R, R'$ and $L$, it
may appear that the last integral cannot be done unless a
semiclassical path is completely specified. However, the system is
integrable, and $[\alpha]$ can in fact be found from the
discontinuities (\ref{pandq}) in terms of $\hat R$. From Eqs.
(\ref{pil}) and (\ref{alpha}), we have
\beq (R'/ L)= \sqrt{1+V}
\cos\alpha, \quad\quad \Pi_L = \pm i G^{-1} R \sqrt{1+V}
\sin\alpha.
\eeq
It follows from (\ref{pandq}) that
\beq
[\sqrt{1+V} \sin\alpha] = \pm i (G p/L R), \quad\quad [\sqrt{1+V}
\cos\alpha] = -(GE/ L R). \label{44}
\eeq
Squaring and adding both
equations in (\ref{44}), we immediately find
\beq \cos[\alpha]
\equiv \cos(\alpha_+ - \alpha_-) = {2+V_++V_--(4\pi G\sigma\hat
R)^2 \over 2 \sqrt{1+V_+}\sqrt{1+V_-}},
\eeq
where the subindices
$+$ and $-$ refer to the limiting values on both sides of the
wall. Note that $V_+ = V(R(q+\epsilon))$ and
$V_-=V(R(q-\epsilon))$ can be different if the cosmological
constants are different on both sides of the wall, or if the
bubble configuration has non-vanishing mass $M\neq 0$. Eq.
(\ref{44}) determines $[\alpha]$ in terms of $\hat R$, but only up
to a sign. Since $0\leq \alpha_\pm \leq \pi$, we have
\beq
{\rm
sign}[\alpha] =-{\rm sign}[\cos\alpha]=-{\rm
sign}[R'/\sqrt{1+V}].\label{signs0}
\eeq
In particular, we shall
see in the Appendix that in the applications we are interested in,
one of these two conditions are met: either we can construct a
semiclassical path for which $p=0$, or we can construct a
semiclassical path for which $V_+=V_-$. In both cases, it follows
that \beq {\rm sign}[\alpha]={\rm sign} E ={\rm sign}\sigma
\label{signs}. \eeq

We are interested in the change $i\Delta W$ between the turning
point geometries $R_b$ and $R_a$,
\beq
\pm\ i \Delta W =
\Omega(R_a)-\Omega(R_b) - \int_{\hat R_b}^{\hat R_a} {{\rm
sign}[\alpha]\over G}\arccos\left[{2+V_++V_--(4\pi G\sigma\hat
R)^2 \over 2 \sqrt{1+V_+}\sqrt{1+V_-}}\right]\hat R d\hat R.
\label{general}
\eeq
where it is now understood that the inverse
cosine lies between $0$ and $\pi$. The value of $\Omega$ at the
turning point geometries is easy to evaluate. The first term in
the integrand of (\ref{omega}) vanishes, and in the second term
$\alpha = \pi \Theta(-R')$, where $\Theta$ is the step function.
As noted in \cite{fmp}, the integral only receives contributions
where the geometry is backtracking $(R'<0)$, and the integral
yields
\beq \Omega = (\pi/2G)\Delta R^2, \label{back}
\eeq
where
$\Delta R^2$ is the absolute value of the change in $R^2$ in the
backtracking part of the geometry.

Let us now particularize the general formula (\ref{general}) to
the problem of nucleation of walls in deSitter space, where the
turning point geometries are given by (\ref{s4}) and (\ref{caps}).
For deSitter space
\beq \Omega(R_b)={\pi \over 2GH^2}.
\eeq
For
the nucleation of positive tension walls, the geometry backtracks
from $R_w$, given by (\ref{rw}), to $R=0$, and so
\beq
\Omega(R_a)={\pi R_w^2 \over 2G}, \quad\quad (\sigma>0).
\eeq
On
the other hand, for the nucleation of negative tension walls, we
have two backtracking pieces. First from $R=H^{-1}$ to $R=R_w$,
and then from $R=H^{-1}$ to $R=0$. Both contribute and give rise
to
\beq
\Omega(R_a)={\pi \over GH^2}-{\pi R_w^2 \over 2G},
\quad\quad (\sigma<0).
\eeq
In both cases, we have \beq
\Delta\Omega = (\Omega_a-\Omega_b) = {\pi\over 2G}(R_w^2 -
H^{-2})\ {\rm sign \sigma}.\label{fst} \eeq The integral in
(\ref{general}) is taken between $\hat R_b=0$, corresponding to a
wall of infinitessimal size before tunneling, and $\hat R_a =
R_w$. Also, since we are discussing the case where the
cosmological constant outside and inside the wall are the same,
and where the initial bubble has infinitessimally small mass, we
have $V_+=V_- = -H^2\hat R^2$. The third term in (\ref{general})
becomes
\begin{eqnarray}
{-{\rm sign}\sigma \over 2G H^2}\int_0^{x_w}
\arccos\left[{1-(1+8\pi^2G^2\sigma^2H^{-2})x \over 1-x}\right] dx
=\nonumber \\ {-\pi \over 4G H^2}[2\pi G\sigma R_w-(2\pi G\sigma
R_w)^2{\rm sign\sigma}],\label{snd}
\end{eqnarray}
where $x_w=H^2
R_w^2=H^2/[H^2+(2\pi G\sigma)^2]$. Adding (\ref{fst}) and
(\ref{snd}), we have
\begin{equation}
\pm i\Delta W = -{\pi^2\sigma R_w\over H^2}=- {1\over 2} I_E
\end{equation}
where $I_E$ is given by Eq.~(\ref{actp}) [or by Eq.~(\ref{actn})]
for a positive [or negative] tension wall.

Eq.~(\ref{general}) applies also to the case of diwalls. In this
case, $V=0$ and the calculation simplifies somewhat. The turning
point before tunneling is given by a flat three dimensional space
$R_b(y)= y$, whereas the turning point after tunneling is given by
Eq.~(\ref{pm}). The first and second terms in the right hand side
of (\ref{general}) can still be calculated from (\ref{back}). The
flat space has no backtracking part, as $R_b$ is monotonic, and
doesn't contribute, whereas for the turning point after tunneling
we have
$$
{\pi \over 2 G} \Delta R^2 = {1\over 8\pi
G^3}(\sigma_1^{-2}-\sigma_2^{-2}).
$$
The third term in the right hand side of (\ref{general}) includes
a separate contribution from each one of the walls, which is
readily calculated and contributes minus one half of the previous
terms. Thus, we find \beq \pm i\Delta W ={1\over 16 \pi
G^3}(\sigma_1^{-2}-\sigma_2^{-2}) =- {1\over 2} I_E, \eeq where
now $I_E$ is given by (\ref{diwact}).

Thus, both for diwalls nucleating in flat space and for walls
(ghost or ordinary) nucleating in deSitter, the WKB "suppression
factor" is given by \beq e^{-2|\Delta W|} = e^{-|I_E|}.
\label{supfac}\eeq In the next Subsection we shall further
elaborate on the relation between the Euclidean action and the
tunneling rates in general.

Note that, since the system is integrable, the precise form of the
semiclassical path has not been used in deriving (\ref{general})
(except in determining the sign of $[\alpha]$, see the discussion
around Eq.~(\ref{signs})). In the Appendix, we construct a WKB
tunneling path for the cases discussed in this subsection. The
trajectories considered are such that the radii of the walls grow
monotonically from an infinitessimally small size to the turning
point size $R_w$. It is interesting to note, in particular, that
for the case of creation of negative tension walls in deSitter,
the deformation of the geometry is smooth and the three volume is
never zero (This is in contrast with the instanton picture where
the deSitter geometry first disappears and then reappears with a
large bubble in it). Note also that, due to the change
in topology, we do not have the possibility of constructing a
semiclassical path for the case of nucleation of negative tension
walls in exactly flat space. Physically, this seems to be in
agreement with the fact that the tunneling suppression for
nucleation of such objects in deSitter becomes insurmountable in
the limit $H\to 0$.

\subsection{The Euclidean action and the
tunneling rates}

Eq.~(\ref{supfac}) suggests that given a semiclassical path with
Euclidean action $I_E$, the corresponding tunneling rate is given
by \beq \Gamma \sim e^{-2|\Delta W|} = e^{-|I_E|}.
\label{tura}\eeq For systems with a single degree of freedom, the
tunneling suppression is indeed given by $e^{-2|\Delta W|}$.
However, for loosely bound systems of ordinary and ghost matter,
the situation is not so clear, and the above formula need not be
of general validity.

To illustrate this point, let us consider the Euclidean solution
corresponding to the nucleation of diwalls in flat space,
considered in Subsection 2.1. We may in fact study the analogous
solution in the presence of a cosmological constant. It is
straightforward to show that the corresponding Euclidean action is
given by \beq I_E^{diwall}(\sigma_1,\sigma_2,H) =
I^+_E(\sigma_1,H)+I^-_E(\sigma_2,H) \eeq where $I_E^\pm$ are the
actions for nucleation of positive and negative domain walls in
deSitter, given by Eqs. (\ref{actp}) and (\ref{actn}). Note that,
since $I_E^+$ and $I_E^-$ have opposite sign, we have \beq
|I_E^{diwall}(\sigma_1,\sigma_2,H)| <
|I^+_E(\sigma_1,H)|+|I^-_E(\sigma_2,H)|. \eeq This means that if
Eq.~(\ref{tura}) is valid, it should be easier to form a diwall,
than it is to create a wall of negative tension and then,
subsequently, another wall of positive tension. Is this a
reasonable expectation?

\begin{figure}%[ht]
%%\vspace{0.5in}
\begin{center}\leavevmode  %
\epsfxsize=15cm \epsfbox{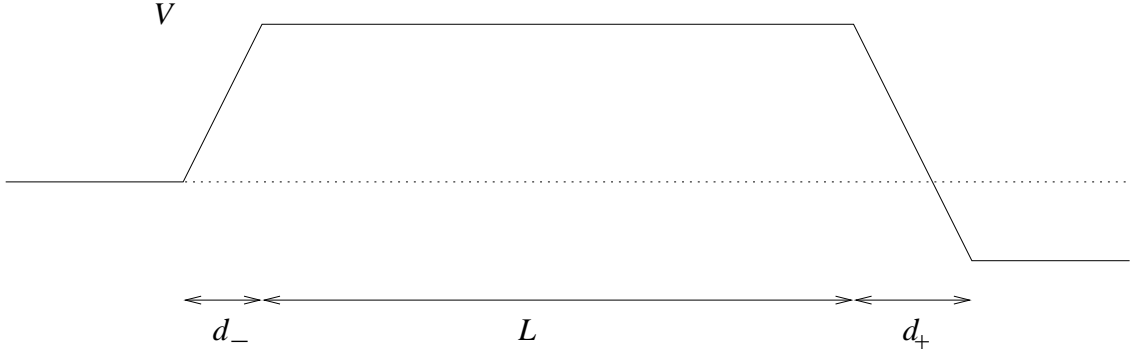}
\end{center}
%%\vspace{-0.5in}
\caption{\small Electric potential barrier exemplifying how a bound
system can behave very differently than its constituents in a tunneling
process. The probability that a proton tunnels through this potential is
negligible for large $L$. An electron feels an inverted potential, and
if its energy is small enough it cannot appear as an asymptotic state in
the region to the right. However, a hydrogen atom can easily cross the
barrier as long as it is not ionized by the electric field. Similarly, a
diwall, or a pair of Bondi dipoles, can be nucleated out of empty space
if the normal and ghost constituents are tightly bound by gravity, even
if their separate nucleation rates are strongly suppressed.
}
\label{fig:potential}
\end{figure}
A simple system wich bears a useful analogy with the system of diwalls
is that of an electron and a proton crossing an electric barrier.
Consider a one-dimensional barrier (fig.~\ref{fig:potential}) where from
left to right, the electric field $-E$ is negative in a segment of width
$d_-$, vanishing in a large segment of length $L$, and positive with
strength $+E$ in a third segment of width $d_+>d_-$. A proton of very
low kinetic energy impinging from the left bounces off the first
segment, repelled by the electric field. The probability for it to go
through the barrier is exponentially small, and practically vanishing in
the limit of large $L$. Likewise, an electron of very low kinetic energy
has no trouble going through the first two segments of the barrier, but
will never make it to the asymptotic region on the right since it simply
doesn't have
enough energy. On the other hand, if the proton and electron are bound
together, the resulting hydrogen atom has no trouble going through the
barrier. The electron pulls the proton through the negative electric
field, and the proton pulls the electron through the third segment with
positive electric field.

Going back to the system of diwalls, let us first consider the limit
$H\ll G\sigma$, when the Hubble radius is much larger than the scale
characterizing the gravitational field of the walls. In this limit, the
walls can be thought as a tightly bound system. In flat space ($H\to 0$)
both $I_E^+$ and $I_E^-$ diverge. Physically, it is impossible to
nucleate either a positive energy wall or a negative energy wall in
Minkowski. Separately, both processes would lead to a breakdown of
energy conservation (barring the possibility of topology change). Yet,
the diwall process seems to be possible and it should occur at a finite
rate. The two walls push and pull each other through a barrier which
none of them would be able to penetrate separately. In this sense, it is
not surprising that Eq.~(\ref{tura}) gives a higher probability to
nucleation of diwalls than to the separate nucleation of walls of either
tension.

On the other hand, in the limit $G\sigma \ll H$ the gravitational field
of the walls (and their interaction) becomes negligible. In this limit,
the walls can be treated as non-gravitating objects in the background
external field of a fixed deSitter space. The instanton for nucleation
of walls of positive or negative tension is just a maximal worldsheet
3-sphere of radius $H^{-1}$ embedded in $S^4$. The corresponding action
is $I_E^{\pm} \approx \pm 2\pi^2 \sigma H^{-3}$. The action for the
diwall is $I^{diwall}_E \approx 2\pi^2 (\sigma_1-\sigma_2)H^{-3}$, and
in the limit when the positive and negative tension walls have
approximately the same modulus, $\delta\sigma/\sigma\ll 1$, we have
$|I_E^{diwall}|\ll |I^+_E|+|I^-_E|$. Eq.~(\ref{tura}) would suggest that
if $\delta \sigma \ll H^3$, diwall production is unsuppressed, even if
$\sigma\gg H^3$. However, this cannot be true, since in the limit we are
considering the interaction between walls is negligible. Production of
walls (and diwalls) should be suppressed, with an exponent which is
parametrically of order $\sigma/H^3$. In the example of the proton and
electron crossing an electric barrier, this limit of weakly bound
components would correspond to the case when there is a thermal bath of
temperature $T\gsim V_0$, where $V_0$ is the ionization energy, or when
the electric field is intense enough to ionize the atom $E r_b \gsim
V_0$, where $r_b$ is the Bohr radius. In both cases, the proton cannot
get hold of the electron, and both have to go through the barrier on
their own, so tunneling is highly suppressed.

The above examples suggest that Eq.~(\ref{tura}) is a good estimate of
the nucleation rate only in the case when we have a tightly bound system
of ghosts and ordinary matter. However, it overestimates the rate when
the gravitational binding energy between ghost and matter components is
weak compared to any other force involved in the tunneling (\eg the
expansion of the universe caused by $\Lambda$ in the case of wall
production in deSitter, the difference in pressure on both sides of the
wall if we consider simultaneous false vacuum decay in ghost and matter
sectors, or the electric force if we consider the Schwinger process
occurring simultaneously in both sectors). In the general case, Eq.
(\ref{tura}) is at best a conservative upper bound to the nucleation
rate.

In the following sections, we shall be interested in nucleation of lumps
of ghost and ordinary matter in flat space. In this case, the system can
be considered to be tightly bound, since the gravitational
energy-momentum transfer between matter and ghosts is what allows the
system to nucleate (there is no other driving force). In the light of
the above discussion, we shall adopt Eq.~(\ref{tura}) as our estimate
for the nucleation rate. Investigation of the general case of weakly
bound systems is left for further research.

\section{Nucleation of Bondi dipoles}
\label{sec:bondidip}

We now turn to discussing the spontaneous nucleation of particle-like
ghosts, accompanied with ordinary particles.

\subsection{Preliminaries}

A spherically symmetric lump of neutral ghost matter of mass $m<0$
coupled to non-ghost gravity creates the gravitational field
\beq\label{onedonkey}
ds^2=-
\left(1+\frac{2G|m|}{\rho}\right)dt^2+\frac{d\rho^2}{\left(1+\frac{2G|m|
}{\rho}\right)} +\rho^2 d\Omega^2\,.
\eeq
This metric exhibits a naked singularity at $\rho=0$. This singularity
is usually regarded as a `valuable' one \cite{garyrob}, \ie one not to
be regularized since it signals a pathological negative-energy spectrum
unbounded below. However, here we intend to explore how deadly this
pathology is. So we assume that \reef{onedonkey} is a valid description
of the gravitational field, though only down to distances of the order of the
gravitational cutoff distance scale $1/\mu$, or, for a light ghost lump
$|m|<\mu$, down to the Compton wavelength $1/|m|$.

\begin{figure}%[ht]
%%\vspace{0.5in}
\begin{center}\leavevmode  %
\epsfxsize=6cm \epsfbox{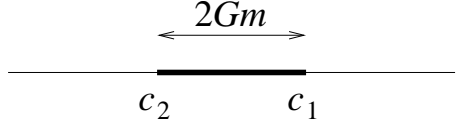}
\end{center}
%%\vspace{-0.5in}
\caption{\small Axisymmetric Weyl solutions are specified by giving the sources
along the axis for the `potential' $U$. The Schwarzschild solution
corresponds to an infinitely thin rod of linear density $+1/2$ and
length $2Gm$.
}
\label{fig:rods1}
\end{figure}

We are interested in configurations with several collinear ghost and
normal particles. The
class of metrics that describes these are the axisymmetric Weyl
solutions
\beq\label{weyl}
ds^2=-e^{2U}dt^2+r^2 e^{-2U}d\phi^2+e^{2(\nu-U)}(dr^2+dz^2)
\eeq
where $U$ and $\nu$ are functions of $r$ and $z$. This is a well-studied
system (see \eg \cite{synge}) so we shall only sketch its solution. The
vacuum Einstein equations can be completely integrated in an explicit
form:
the function $U$ satisfies a Laplace equation in the auxiliary space
$dr^2+dz^2+r^2 d\phi^2$, and then $\nu$ can be obtained from $U$ by
quadratures. Hence the solutions are fully determined once we specify
the sources for the axisymmetric potential $U$. For example, the
Schwarzschild solution with positive mass $m$ corresponds to taking as a
source an infinitely thin rod of length $2Gm$ and linear mass density
$1/2$ along the $z$ axis (fig.~\ref{fig:rods1}), so
\beq\label{schwU}
e^{2U}=\frac{X_2}{X_1}
\eeq
and
\beq\label{schwnu}
e^{2(\nu-U)}=\frac{Y_{12}}{4 R_1 R_2}\,,
\eeq
where we introduce notation that is useful for this class of solutions \cite{fay},
\beqa
R_i&=&\sqrt{r^2+(z-c_i)^2}\,,\\
X_i&=&R_i-(z-c_i)\,,\\
Y_{ij}&=& R_i R_j +(z-c_i)(z-c_j)+r^2\,,
\eeqa
and, for this particular solution, $c_1-c_2=2 G m$.
It may seem odd that the spherically symmetric Schwarzschild solution is
obtained from a non-spherical source, but this is just an artifact of
the Weyl representation. The rod at $r=0$, $c_2<z<c_1$, in fact
corresponds to the Schwarzschild horizon, and it is possible to see in
general that a regular horizon is present iff the rod density is $1/2$,
otherwise a naked singularity will appear. In this paper, however,
we will only consider situations where the geometry is cutoff before the
horizon (or the naked singularity) is reached.

The field of a ghost particle \reef{onedonkey} is given by a
configuration similar to the one above, but now the density of the rod
is instead $-1/2$. A simple way to obtain this from the solution
\reef{schwU}-\reef{schwnu} is to take $c_1<c_2$, with $c_2-c_1=2G|m|$.
For definiteness and simplicity we shall only consider normal and ghost
particles with rod linear densities $\pm 1/2$ ---other densities do not
introduce any significant
qualitative changes.
\begin{figure}%[ht]
%%\vspace{0.5in}
\begin{center}\leavevmode  %
\epsfbox{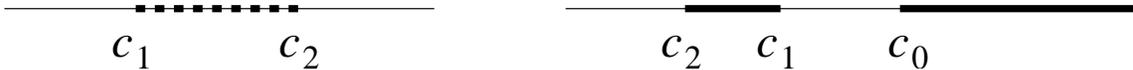}
\end{center}
%%\vspace{-0.5in}
\caption{\small Left: the source for a ghost particle of mass $-m_-$ is
a rod of length $2Gm_-$ and linear density $-1/2$ (dotted rod). This
can be obtained from eqs.~\reef{schwU}-\reef{schwnu} by simply taking
$c_1<c_2$. Right: the C-metric describing accelerating black holes
corresponds to a finite rod for the black hole, and a semi-infinite rod
for the acceleration horizon. Weyl coordinates cover only one Rindler
wedge of the entire spacetime.
}
\label{fig:rods2}
\end{figure}

It is now easy to construct axisymmetric configurations with arbitrarily
many collinear particles. One simply superposes rods corresponding to
each of the particles. $U$ is then the linear superposition of the
fields of each rod, and then the function $\nu$ can be explicitly
integrated \cite{iskh,fay,er}. Of course, in general the particles will
attract each other (or possibly repel if ghosts are present) so there
will be uncancelled forces among the particles. This reflects itself in
the geometry through the presence of conical singularities on the
portions of the $z$-axis away from the rods. For Weyl metrics
\reef{weyl}, the conical deficit angle $\delta$ at a given point $z_0$
on the axis away from any
rod, is given by
\beq\label{condef}
2\pi-\delta=\lim_{r\to
0}\frac{2\pi}{\sqrt{g_{rr}}}\frac{d\sqrt{g_{\phi\phi}}}{dr}|_{z=z_0}=
2\pi\lim_{r\to
0} e^{-\nu}|_{z=z_0}\,.
\eeq
A conical {\em deficit} angle $\delta>0$ on a certain segment of the
axis can be interpreted as a string stretched along the segment and
pulling together the objects at its endpoints, while a conical {\em
excess} angle $\delta<0$ is a strut pushing the objects apart. An
integration constant in $\nu$ (corresponding to constant rescalings of
$\phi$) can be used to eliminate possible conical singularities on a
given segment, but in general, once this is fixed, there will remain
conical singularities at other segments. In some cases it will be
possible to tune parameters in the solution to cancel all conical
singularities: the system is then balanced. If it is impossible to
cancel all of them, then external forces are needed to keep the
configuration in place.

Finally, we note that a semi-infinite rod of density $+1/2$ corresponds
to an acceleration horizon. The Weyl coordinates cover in this case only
a Rindler wedge of the whole spacetime, but they can be extended across
the Rindler horizon to provide a maximal analytic extension of the
geometry, which contains a second copy of the Rindler wedge
\cite{iskh,bonnor}. For instance, the configuration corresponding to a
finite, positive mass rod, and a semi-infinite rod, is known as the
C-metric (see fig.~\ref{fig:rods2}) and describes two black holes
accelerating away with uniform
acceleration, each in a separate Rindler wedge, and being pulled apart
by strings \cite{kw,bonnorC}.

\subsection{Accelerating Bondi dipoles}

\begin{figure}%[ht]
%%\vspace{0.5in}
\begin{center}\leavevmode  %
\epsfxsize=8cm
\epsfbox{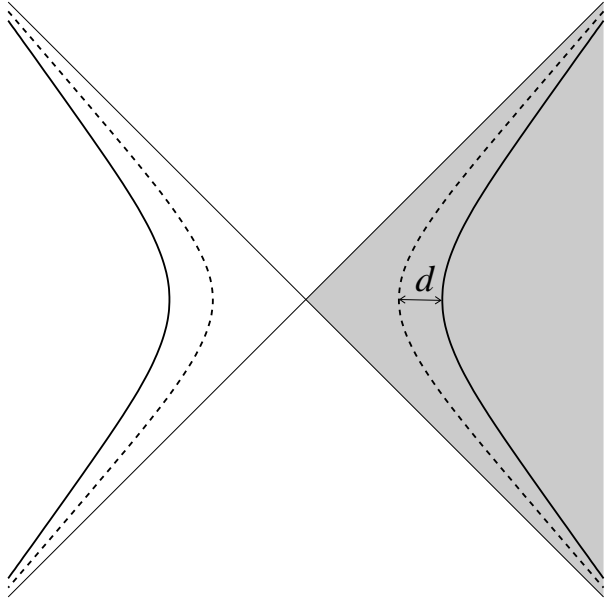}
\end{center}
%%\vspace{-0.5in}
\caption{\small A pair of Bondi dipoles accelerating apart. Weyl
coordinates cover only one wedge (shaded) of the full spacetime,
containing a single Bondi dipole. A normal particle (solid worldline) is
chased by a ghost (dotted worldline). The dipoles self-accelerate away
with uniform acceleration, and the configuration is boost-invariant.}
\label{fig:bondi}
\end{figure}
A Bondi dipole consists of a positive and a negative mass particle
\cite{bondi}. If we construct the gravitational field of this
configuration as the Weyl metric for a positive density and a negative
density rod, and nothing else, then it is straightforward to see that no
choice of the rod parameters is able to cancel all conical singularities
on all segments of the axis away from the rods. In particular, if we
eliminate the conical defects at infinity, then a strut remains
inbetween the two particles. This is of course expected, since the two
particles can not remain in static equilibrium. The strut pushes the
normal particle away from the ghost, while the ghost is also pushed
away, but having negative mass, it will tend to accelerate in a
direction opposite to the one it is pushed in\footnote{Hence Gamow's
term ``donkey particle'' for such ghosts \cite{gamow}.}. This indicates
that the Bondi dipole will naturally accelerate together, the ghost
chasing after the normal particle in a runaway fashion.

\begin{figure}[ht]
%%\vspace{0.5in}
\begin{center}\leavevmode  %
\epsfbox{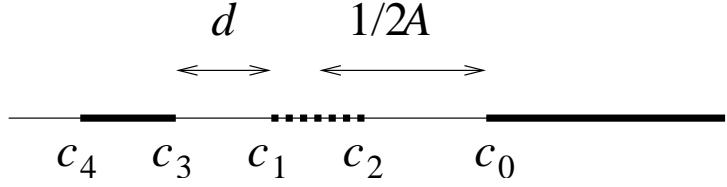}
\end{center}
%%\vspace{-0.5in}
\caption{\small Weyl sources for an accelerating Bondi dipole: a ghost
particle (negative density rod at $c_1<z<c_2$) chases a normal particle
(positive density rod at $c_4<z<c_3$). The maximally extended solution
contains a second pair accelerating in the opposite direction.
}
\label{fig:rods3}
\end{figure}
{}From our previous discussion, it should be obvious how to construct
the Weyl metric for such a configuration\footnote{A slightly different
version of the Bondi dipole was constructed in \cite{bonnor}, with
point-like sources for the normal and ghost particle. These are then
``Chazy-Curzon particles". For our purposes, the qualitative properties
are basically the same as in our solution. In particular, the result for
the action of the corresponding instanton takes the same form as in
eq.~\reef{bondiact}. Other instances of self-accelerating
positive+negative mass bound states are discussed in \cite{zeromass}.}.
From left to right, take the sources for the potential $U$ to be (figure
\ref{fig:rods3}): a semi-infinite rod at $z>c_0$ (and of density $+1/2$)
for the acceleration horizon; a negative density rod at $c_1<z<c_2<c_0$
for the ghost; and a positive density rod at $c_4<z<c_3<c_1$ for the
normal particle. Note that again we have chosen to label the rod
endpoints $c_i$ in such a way that the conventional correlative order is
inverted for the negative density rod, \ie we take $c_1<c_2$. This is
done in order to facilitate a direct
comparison to the solution in \cite{fay}, which described two normal
particles (black holes) accelerating together. Then we can directly
read off our solution from \cite{fay} as
\beqa\label{bondipair}
e^{2U}&=&A_-\frac{X_0 X_2 X_4}{X_1 X_3}\,,\nonumber\\
e^{2(\nu-U)}&=& \frac{Y_{01}Y_{12}Y_{03}Y_{34}Y_{14}Y_{23}}{8 A_- R_0 R_1
R_2 R_3 R_4 Y_{02}Y_{04}Y_{13}Y_{24}}
\eeqa
with
\beq\label{order}
c_4<c_3<c_1<c_2<c_0\,,
\eeq
and $A_-$ a constant with dimension of inverse length whose value can be
changed by coordinate rescalings. For later
convenience we choose it as
\beq\label{ac}
A_-^{-1}=2c_0-c_1-c_2\,.
\eeq

The inversion in the ordering between $c_1$ and $c_2$ in \reef{order}
compared to \cite{fay} changes a positive mass into a negative one. Note
also that we can always perform a shift in $z$ to fix one of the $c_i$,
so the solution contains only four physical parameters. To identify the
physical quantities, we take the approximation where the rods are well
separated from each other, and the dipole has a size smaller than the
acceleration length
scale, \ie
\beq\label{rodapart1}
c_3-c_4 \sim c_2-c_1\ll c_1-c_3 \ll c_0-c_2\,,
\eeq
so the gravitational forces between the particles involved are weak. A
careful examination of the solution in different limits then allows to
identify\footnote{These identifications are not uniquely
determined: they can be modified by terms that are small in the limit
\reef{rodapart}. Our choices simplify some expressions below.}
\beqa\label{masses}
m_+&=&\frac{c_3-c_4}{2G}\frac{A_+}{A_-}\,,\qquad m_-=\frac{c_2-c_1}{2G}\,,
\\
A_+^{-1}&=&\sqrt{\frac{2c_0-c_3-c_4}{A_-}}\,,\qquad d=c_1-c_3\,.\nonumber
\eeqa
All these quantities are invariant under shifts along the $z$ axis. The
approximation \reef{rodapart} is now equivalent to
\beq\label{rodapart}
G m_+\sim G m_-\ll d\ll A_-^{-1}\sim A_+^{-1}\,,
\eeq
and in this limit we can approximately interpret $m_+$, $A_+$ and
$-m_-$, $A_-$, respectively, as the masses and accelerations of the
positive and negative particles, and $d$ as the distance between them.
In particular, in the approximation \reef{rodapart} the harmonic
difference of accelerations satisfies (see
figure \ref{fig:bondi})
\beq\label{dA}
\frac{1}{A_+}-\frac{1}{A_-}\simeq d\,.
\eeq

Note that with our definition, $m_-$ is positive, so the ghost mass is
$-m_-$. We can take the four physical parameters of the solution to be
$m_\pm$, $d$ and $A_-$, while $A_+$ is fixed in terms of them.

In \reef{bondipair} we have already fixed the integration
constant in $\nu$ so as to have
\beq\label{cone1}
\lim_{r\to 0} e^{-\nu}|_{(z<c_4)}=1
\eeq
and therefore, according to \reef{condef}, there is no conical defect at
infinity as long
as $\phi\sim\phi+2\pi$. On the other hand, at the segment inbetween the
two rods we have
\beq\label{cone2}
\lim_{r\to 0}
e^{-\nu}|_{(c_3<z<c_1)}=\frac{(c_0-c_3)(c_2-c_3)(c_1-c_4)}{(c_1-c_3)(c_0-
c_4)(c_2-c_4)}
\eeq
and between the ghost rod and the acceleration horizon,
\beq\label{cone3}
\lim_{r\to 0}
e^{-\nu}|_{(c_2<z<c_0)}=\frac{(c_0-c_1)(c_0-c_3)}{(c_0-c_2)(c_0-c_4)}\,.
\eeq
To cancel conical singularities on all segments we must require
equality of \reef{cone1}, \reef{cone2}, \reef{cone3}, \ie
\beq\label{nocone}
\frac{(c_2-c_3)(c_1-c_4)}{(c_1-c_3)(c_2-c_4)}=\frac{c_0-c_1}{c_0-c_2}
=\frac{c_0-c_4}{c_0-c_3}\,.
\eeq
The conditions \reef{nocone} leave only
two free parameters in the solution, which we could take to be the two
masses $m_+$, $m_-$, of the dipole constituents. In that case, the size
and acceleration of the dipole would be fixed.

After some
algebra and using the definitions \reef{masses}, the second
equality in \reef{nocone} can be exactly recast into the form of Newton's third law
\beq\label{newt1}
m_+A_+=m_-A_-\,.
\eeq
Since necessarily we have $A_+<A_-$, this implies that
\beq\label{massdiff}
m_+>m_-\,.
\eeq
This requirement will be important later on.

Also, in the approximation of weak fields \reef{rodapart}, the conditions
\reef{nocone} become
\beq\label{newt}
\frac{Gm_+m_-}{d^2}\simeq m_-A_-=m_+A_-\,,
\eeq
which reproduce Newton's force law, as expected.
Finally, in this approximation eqs.~\reef{nocone}
imply
\beq\label{weakd}
m_+ -m_-\equiv \Delta m\simeq\frac{G m_+m_-}{d}\,.
\eeq
Hence the relative
difference between the (absolute values of the) masses of the two
components of the dipole is small relative to their masses.

\subsection{Pair creation of Bondi dipoles}

If we analytically continue the solution of the previous subsection to
imaginary time, we obtain a Euclidean instanton where the dipole runs in
a loop, with Euclidean time playing the role of the angle coordinate in
the loop. Since the solution is asymptotically
flat, the initial state is the Minkowski vacuum, which then tunnels to a
zero-energy configuration with two Bondi dipoles self-accelerating away
in opposite directions. This decay of the Minkowski vacuum is the
non-perturbative analogue of the process \reef{decay1},

In order to compute the Euclidean action $I_E$, we follow \cite{HaHo}.
The periodicity of Euclidean time is fixed by requiring regularity at
the Rindler horizon only ---we are assuming that the normal particle has
no horizon. As long as we assume that the region near the ghosts is
smoothed out by ghost matter fields that satisfy the classical Einstein
equations, we do not need to know the details of how the naked
singularity is avoided---nor how normal matter modifies the geometry to
avoid the black hole horizon. Bulk terms in the action (including matter
and ghost fields) vanish on-shell, and only boundary terms contribute.
These are entirely given in terms of the gravitational field. The action
contains a contribution from the Hamiltonian,
which vanishes for this zero-energy configuration, plus terms arising
from the horizons. We are assuming that the positive mass particles
possess no black hole horizons, so the only contributions come from the
acceleration horizons. The action is then
\beq\label{actionarea}
I_E=-\frac{\Delta\mathcal{A}}{4G}
\eeq
where $\Delta\mathcal{A}=\mathcal{A}^{(f)}-\mathcal{A}^{(i)}$ is the difference
between the areas of the acceleration horizons after and before the
nucleation process, \ie in the instanton and in the reference Minkowski
background. Each of these is evaluated as
\beq
\mathcal{A}=\int d\phi \int_{c_0}^{z_\mathrm{max}} dz
\sqrt{g_{zz}g_{\phi\phi}}|_{r=0}\,.
\eeq
Since the
area of acceleration horizons is infinite, we have regularized
them with a long distance cutoff $z_\mathrm{max}\gg c_0$. In general
$z_\mathrm{max}^{(f)}\neq z_\mathrm{max}^{(i)}$, and in order to make sure
that the subtraction is correctly performed, one must match the lengths
of the two acceleration horizons
\beq
\ell=\int d\phi \sqrt{g_{\phi\phi}}|_{r=0, z=z_\mathrm{max}}\,.
\eeq
Imposing $\ell^{(f)}=\ell^{(i)}$ fixes the relation
between the long-distance cutoffs $z_\mathrm{max}^{(i)}$ and
$z_\mathrm{max}^{(f)}$. For finite cutoffs, the two areas are
finite and we can
compute their difference. Finally, removing the regulators leaves a finite
non-zero result
\beq\label{bondiact}
I_E=\frac{\pi}{2GA_-}(c_1-c_2+c_3-c_4)=\pi\left(\frac{m_+}{A_+}-\frac{m_-}{A_-}\right)\,.
\eeq
As an aside, we note that this result can be easily generalized to
configurations with arbitrarily many accelerating particles and ghosts.
If the finite rods (of density $\pm 1/2$) have endpoints at $z=c_i$, then the contribution to
the action coming from $\Delta\mathcal{A}$ is of the form
$\frac{\pi}{2AG}\sum_i\pm c_i$, where the sign for $c_i$ is positive (negative) if
the density along the axis increases (decreases) as we go
from right to left at $z=c_i$. If we identify the masses and
accelerations with appropriate generalizations of \reef{masses}, then this
contribution to
the action is equal to $\pi \sum_i \pm \frac{m_i}{A_i}$, where ghosts
contribute with a minus sign.

Of course, the requirement of balance of forces (\ie absence of
conical singularities) imposes relations between the parameters.
Coming back to our example of the Bondi dipole, the condition
\reef{massdiff} together with $A_+<A_-$ imply that the action
\reef{bondiact} is always positive and the decay rate is
exponentially suppressed. In the weak field approximation the
result reduces to
\beq\label{weak}
I_E\simeq 2 \pi m_+ d\ \simeq
{\Delta m \over T_R},
\eeq
where the
Rindler temperature is $T_R=A_+/2\pi$. The rate of the
non-perturbative version of the decay \reef{decay1} is therefore
\beq \Gamma\sim e^{-2\pi m_+ d}\,. \eeq

\section{Materialization of ghost couples}
\label{sec:ghostpairs}

\subsection{Ghost pairs in Minkowski space}

Can we also find a non-perturbative analogue of the decay process
\reef{decay2}, in which no normal particles are produced? Since
\reef{decay2} involves the creation of gravitons,
non-perturbatively the decay might not only produce gravitational
waves but also change significantly the final geometry. We shall show
that this is indeed the case, and that the Minkowski vacuum undergoes a
topology change.

We expect the solution that describes the decay to involve two ghost
particles accelerating away from each other. In the terminology of the
Weyl solutions discussed above, this should correspond to a
semi-infinite rod for the acceleration horizon, and a finite rod of
negative density for the ghost particle. From our Newtonian estimates in
the introduction section, the two particles should be very close, $r\sim
Gm_-$, \ie very close to the acceleration horizon. Therefore a strong
gravitational backreaction is expected.

A simple way to construct this solution is to start from the configuration in
the previous section (eq.~\reef{bondipair} and fig.~\ref{fig:rods3}) and
remove from it the normal particle, \ie the positive density rod, by
setting $c_3=c_4$.  This
leaves only the ghost particle in the Rindler wedge covered by these
coordinates (its anti-ghost is in the opposite wedge). We must examine,
though, the condition for balance of forces in this new configuration.
With our choice of parameters no conical singularity is present at
$z<c_1$ (\ie out to infinity), but, if $c_3=c_4$, it is
impossible to eliminate the conical defect \reef{cone3} at
$c_2<z<c_0$. The way around this, however, is to take $c_2\to c_0$,
hence eliminating the troublesome segment of the axis. In terms of the
`physical' parameters \reef{ac}, \reef{masses}, this implies
\beq\label{donkeymass}
2Gm_- A =1\,.
\eeq
(from now on we drop the subindex $-$ from $A_-\to A$). Clearly, this means we
are in a regime of strong self-gravity, so the interpretation of $m_-$
as the ghost mass should not be taken any strictly.
\begin{figure}%[ht]
%%\vspace{0.5in}
\begin{center}\leavevmode  %
\epsfbox{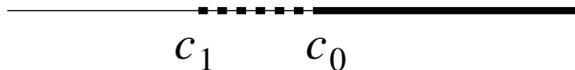}
\end{center}
%%\vspace{-0.5in}
\caption{\small Weyl sources for ghost and anti-ghost accelerating apart
(the anti-ghost lies in the opposite Rindler wedge obtained by maximally
extending the solution).
}
\label{fig:rods4}
\end{figure}

\begin{figure}%[ht]
%%\vspace{0.5in}
\begin{center}\leavevmode  %
\epsfxsize=10cm \epsfbox{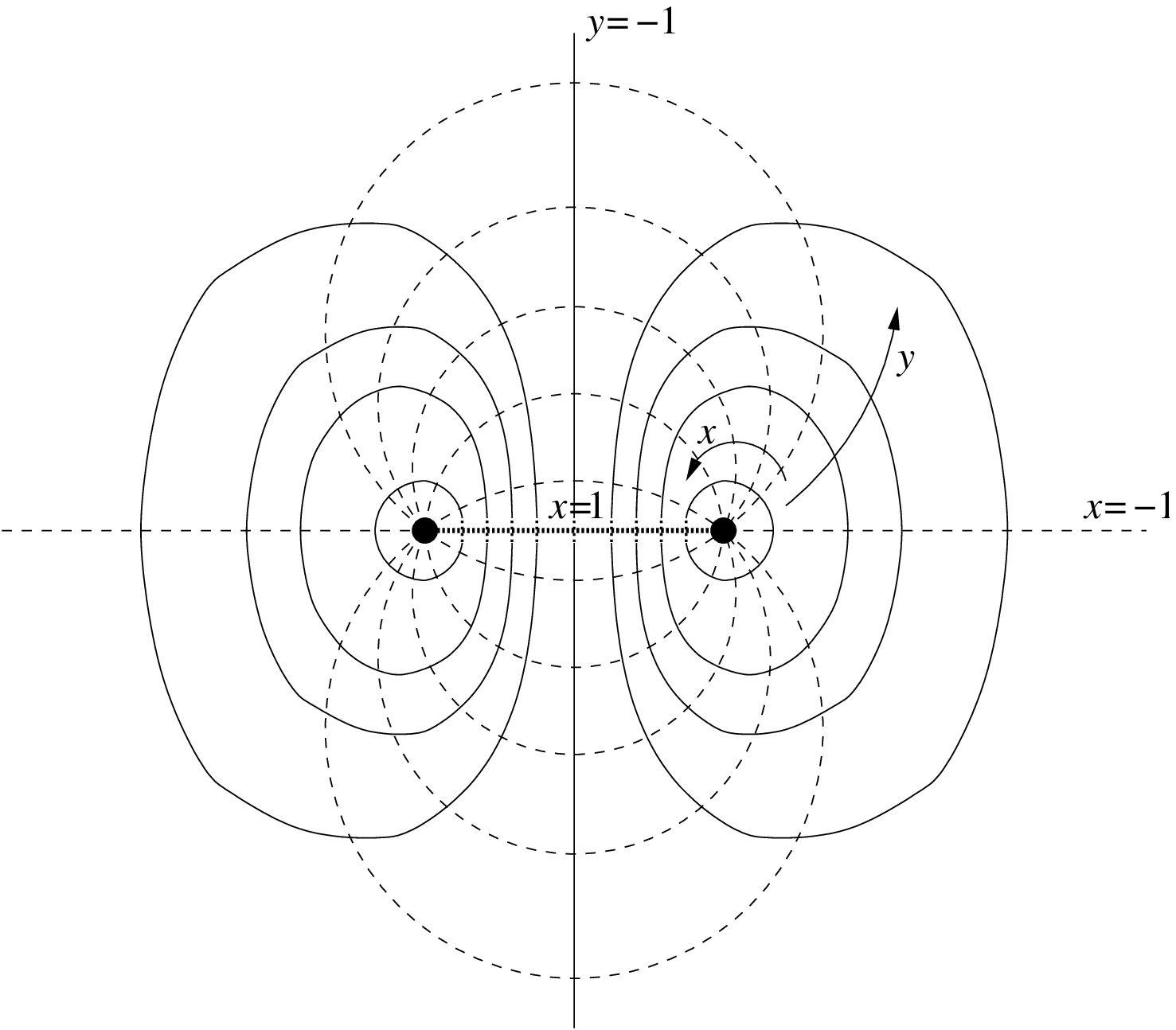}
\bigskip

\epsfxsize=12cm \epsfbox{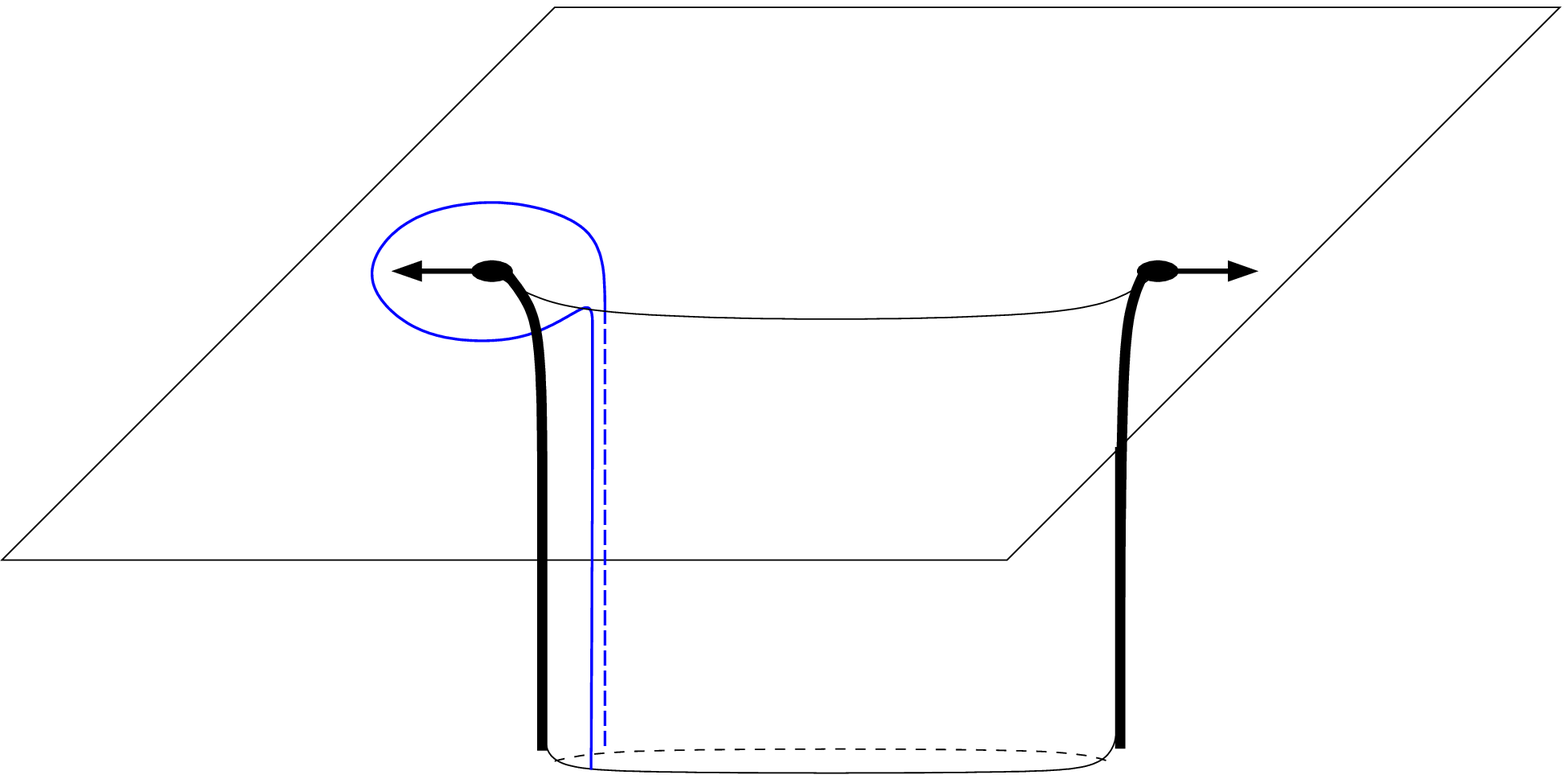}
\end{center}
%%\vspace{-0.5in}
\caption{\small Section at $t=0$ (with the angular direction $\phi$
suppressed) of the geometry with a pair of ghost particles. The upper
figure sketches the coordinate system: solid lines are
constant $y\in(-\infty,1]$, dashed lines are constant $x\in[-1,1)$. The
lines (actually surfaces) at constant $y$, which surround the ghost
singularity at $y=-\infty$, do not quite close at the axis $x=1$, which lies
an infinite spatial distance away. This corresponds to the infinite
throat that stretches between the particles, sketched in the lower
figure by taking a slanted view. A surface at constant $y$ (depicted as
a line) surrounding the ghost can be described as a sphere with an infinite
funnel at one of its poles (so it is topologically ${\mathbb R}^2$, instead
of $S^2$). Since
the ghosts are lumps that spread out to some finite $y$ away from the
singularity, they actually extend all the way down the throat. The
surface $y=-1$ is an acceleration horizon, so the two particles are
causally separated.
}
\label{fig:ghostpair}
\end{figure}

The Weyl form is rather unwieldy if one wants to study in detail the
properties of this configuration. To get a better understanding of the
geometry, it is more convenient to note that this solution is in fact a
special case of the neutral C-metric of \cite{kw} (figure
\ref{fig:rods2}-right), namely, one with a particular negative value for
the mass parameter. We then use a more conventional set of coordinates
(see fig.~\ref{fig:ghostpair}-upper), in which the solution becomes
\beq\label{negC}
ds^2=\frac{1}{A^2(x-y)^2}\left[ G(y)d\hat{t}^2+(1+2Gm_-A)^2\left(
-\frac{dy^2}{G(y)}+\frac{dx^2}{G(x)}\right)+G(x)d{\phi}^2\right]\,,
\eeq
where we use the form for the cubic function $G(\xi)$ advocated in \cite{ht},
\beq
G(\xi)=(1-\xi^2)(1-2Gm_-A\xi)\,.
\eeq
The change of coordinates between $(r,z)$ and
$(x,y)$ can be found in \cite{ht}.
The time coordinate $t$ used in the Weyl form of the metric has, for
convenience, been
rescaled to $\hat{t}=t\frac{A}{1+2Gm_-A}$. Note we
have not imposed the condition \reef{donkeymass} yet. It will
be rederived shortly.

The coordinate range is
$-\infty<y<-1$ and $-1\leq x\leq 1$, with $y\to x\to -1$ corresponding
to asymptotic infinity. The acceleration horizon lies at $y=-1$.

For our purposes here, the axes of $\phi$, which lie at $x=\pm 1$, are
of paramount importance. The surfaces at
constant $\hat{t}$ and constant $y>-\infty$, are, up to a conformal
factor, described by
\beq\label{twosphere}
(1+2Gm_-A)^2\frac{dx^2}{G(x)}+G(x)d\phi^2\,.
\eeq
As long as $2Gm_-A<1$, the function $G(x)$ has simple zeroes at $x=\pm
1$. The coordinate $x$ can be regarded as roughly similar to
$\cos\theta$ on $S^2$, so these surfaces are topological two-spheres
enclosing the ghost particle, and $x=\pm 1$ are the poles of the
spheres. It is easy to see that if we identify $\phi\sim \phi+2\pi$,
then there is no conical defect at $x=-1$, \ie extending out to
infinity. There is, though, a conical excess at the other pole $x=+1$,
which stretches from the ghost towards the acceleration horizon, and
thence to the anti-ghost. It is easy to see that this conical excess
cannot be eliminated by any choice of parameters --- just like we
already found using the Weyl form of the solution. However, if we impose
the condition \reef{donkeymass}, then the topology of these surfaces
changes. Now the function
\beq
G(x)=(1-x)^2(1+x)
\eeq
has a double zero at $x=1$. As a result, the singularity at $x=1$ is
pushed an infinite spatial distance away, so it is removed from the
spacetime. The coordinates $(x,\phi)$ parametrize now a surface with
topology $\mathbb{R}^2$, instead of $S^2$. Geometrically it is a sphere
with an infinite funnel at one of its poles (figure
\ref{fig:ghostpair}-lower). Down the funnel at $x\to
+1$ it is convenient to introduce a new coordinate
\beq\label{funnelz}
\zeta=-\log (1-x)
\eeq
in terms of which the geometry \reef{twosphere}, when the condition
\reef{donkeymass} holds, approaches
\beq\label{funnnel}
2\left(d\zeta^2+e^{-2\zeta}d\phi^2\right)
\eeq
with $\zeta\to \infty$: the tip of the funnel recedes to infinity as
its width shrinks.

Thus the effect of choosing the parameters to satisfy \reef{donkeymass}
is to create an infinite throat
that extends between the ghost and anti-ghost. Away from the naked
singularities at the position of the ghosts ($y\to -\infty$), the
geometry is regular.
Down the throat, when $x\sim
+1$, and within the Rindler wedge $-\infty<y\leq -1$, the solution can
be approximated by
\beq\label{throat}
ds^2\simeq\frac{1}{A^2(1-y)^2}\left[ G(y)d\hat{t}^2
-4\frac{dy^2}{G(y)}+2\left(d\zeta^2+e^{-2\zeta}d\phi^2\right)\right]\,.
\eeq
The size of $\phi$-circles at constant $x$ goes from zero at the
position of the ghost, $y\to -\infty$, to a maximum at the midpoint $y=-
1$.

Figure~\ref{fig:ghostpair} also reflects one peculiarity of this
construction. Even though we have been referring to the singularities in
this metric as ghost particles, the fact is that they are qualitatively
different from the ones in \reef{onedonkey}. In particular, and in
constrast to the ghosts in the Bondi dipole of the previous section,
there is no limit of this solution where one recovers the single ghost
solution \reef{onedonkey}. Then it becomes doubtful whether the ghost
singularities in this solution, which are more string-like than
particle-like, can be identified as localized ``lumps of ghost matter".
This caveat should be borne in mind throughout our discussion.

As the ghost and antighost accelerate away, the infinite throat
inbetween them grows. We can study the geometry at asymptotically late
times using methods similar to those employed in \cite{emgu}. First,
note that asymptotic future lies in the upper wedge of the spacetime.
To describe this region, we take $y>-1$, \ie we cross the Rindler
horizon towards the future, extending the solution in the standard
manner. Since in this region $G(y)>0$, we now have $y$ as the timelike
coordinate, while $t$ is spacelike. This means that the geometry in this
region is time-dependent.

The range of coordinates that describes this ``upper
wedge" portion of spacetime is
$-1\leq y<x\leq 1$. There is no singularity in this region. Asymptotic
future is approached as $y\to x\to 1$ with $y<x$. In this limit the
throat stretches out to infinity as the two particles fly away. To
describe the resulting geometry when the ghosts are far away, consider
$1-x\ll x-y\simeq 1-y$, and introduce, besides \reef{funnelz}, new
coordinates that measure
proper distance and time
\beq\label{tomilne}
w=\frac{\sqrt{2}}{A}\hat{t}\,,\qquad\tau=\frac{\sqrt{2}}{A(1-y)}
\eeq
(recall $\hat{t}$ is spacelike and $y$ is timelike in the upper
wedge).
The spacetime in the aftermath of the decay is then
\beq\label{aftermath}
ds^2\to
-d\tau^2+\tau^2\left(d\zeta^2+e^{-2\zeta}d\phi^2\right)
+dw^2\,\qquad(\mathrm{asymptotic\; future}).
\eeq
It is now apparent that this is flat space: besides the flat coordinate
$w$ (which was Rindler time $\hat{t}$ in the Rindler wedge) the rest of
the spacetime is the three-dimensional Milne universe, which is locally
flat. The Milne universe is an open ($k=-1$) FRW universe with spatial
sections of constant negative curvature. Note, however, that in the
spatial sections in \reef{aftermath} there is no need \textit{a priori}
to impose any periodicity requirements on $\phi$, since $\partial_\phi$
does not have any fixed-points at finite spatial distance. But, since
this is an extension of the solution from the Rindler wedge, we
\textit{must} have the same periodicity for $\phi$ on both sides of the
Rindler horizon, \ie $\phi\sim\phi +2\pi$ also in \reef{aftermath}.
Although locally the geometry is unaffected, these identifications change
the spatial sections of the Milne universe from a hyperboloid of
topology ${\mathbb R}^2$ to an infinite funnel of topology ${\mathbb R}\times
S^1$, the former being the universal covering of the latter.

To summarize, the non-perturbative production of a pair of ghosts makes
flat spacetime undergo a topology change from ${\mathbb R}^{1,3}$ to
${\mathbb R}^{1,2} \times S^1$. In the future, when the ghosts are far
away, one finds an asymptotically flat space in which a (non-singular)
string-like object has formed. The `core' of the string is an infinite
throat where a spatial direction is compactified into a circle of
shrinking radius. The geometry is time-dependent, and presumably
contains Einstein-Rosen cylindrical gravitational waves produced in the
nucleation event.

In order to compute the Euclidean action for the instanton we shall use the
results in the previous section. The appearance of a new
asymptotic region down the throat does not give any additional
contribution to the action of \reef{bondiact}: one can easily check that
the Gibbons-Hawking boundary term $\int\sqrt{\gamma}\;K$ vanishes at the
boundary at $x=+1$. Hence we can directly read off the result from
eq.~\reef{bondiact}, after making the parameter choices appropriate for this
solution. We find
\beq\label{donkeyact}
I_E=-\frac{\pi}{2GA}(c_0-c_1)=-\frac{\pi m_-}{A}=-2\pi G m_-^2\,.
\eeq
This action is {\it negative}. However, as we have explained in section
\ref{sec:ghostwalls} above, this does not imply an exponentially
enhanced catastrophic decay. Rather, the production rate is always
suppressed, and is
\beq
\Gamma\sim e^{-|I_E|}=e^{-2\pi G m_-^2}\,.
\eeq

\subsection{Conjuring up ghosts in deSitter space}

In the previous subsection we have seen that the materialization of a
pair of ghosts (without accompanying normal particles) results in the
formation of a new asymptotic region in the shape of an infinite throat
inbetween the ghosts, with a compactified spatial circle. The creation
of such a new infinity may sound a little disturbing and may cast some
doubt on the whole construction. We saw in sec.~\ref{sec:ghostwalls}
that ghost domain wall formation in flat space also came associated with
the appearance of a new asymptotic region. In that case, we found that
the inclusion of a cosmological constant served to regularize the
possible divergences of the infinite volume that arises after
nucleation.

This suggests that we consider the decay of empty deSitter space via the
materialization of two ghosts. Even if, as we will see, the appearance
of an infinite throat is {\em not} avoided in this way, the issues of
regularization of the Euclidean action are much simpler, since the
Euclidean action of the instanton is already finite. The space left
over as the ghosts fly apart is again deSitter, although, as was the
case with the flat space at asymptotic future in the previous section,
it has non-trivial global identifications.

The solution we seek is obtained from a generalization of the C-metric
\reef{negC} to include a cosmological constant $\Lambda=3H^2$ \cite{pd}.
Its metric can be written as
\beq\label{negCdS}
ds^2=\frac{1}{A^2(x-y)^2}\left[ F(y)d\hat{t}^2+(1+2Gm_-A)^2\left(
-\frac{dy^2}{F(y)}+\frac{dx^2}{G(x)}\right)+G(x)d{\phi}^2\right]\,,
%ds^2=\frac{1}{\bar{A}^2(x-y)^2}\left( F(y)d\bar{t}^2
%-\frac{dy^2}{F(y)}+\frac{dx^2}{G(x)}+G(x)d\bar{\phi}^2\right)\,,
\eeq
where
\beq
G(x)=(1-x^2)(1-2Gm_-Ax)\,,\qquad
F(y)=h^2+(1-y^2)(1-2Gm_-Ay)\,,
\eeq
and, to reduce clutter, we have defined
\beq
h= H\frac{1+2Gm_-A}{A}\,.
\eeq
Again, we have already chosen the sign of the mass parameter (with $m_->0$) to
correspond to a ghost. It is
obvious that as $H\to
0$ we recover \reef{negC}. The main difference with the latter is that
the roots of $G(x)$ and $F(y)$ now do not coincide. As a consequence,
while $x$ still varies in the range $[-1,1]$, now $y\in [-\infty,y_0]$,
where $y_0$ is the only real root that $F(y)$ has when
$m_-A>0$ and $H^2>0$. Since $y_0<-1$,
the asymptotics of space differ from \reef{negC}: $y=y_0$ is the
deSitter horizon, and no spatial infinity can be reached (we will see
that $y\to x$ is future infinity).
As before, $y=-\infty$ is the naked singularity of the ghost.

The $(x,\phi)$ sector in \reef{negCdS} is exactly the same as in \reef{negC}, so
the analysis of the geometry of \reef{twosphere} in the previous section
applies again: we remove the conical singularity at $x=-1$ by
identifying $\phi\sim\phi+2\pi$. To
eliminate the singularity at $x=1$, the only choice is
eq.~\reef{donkeymass}, which pushes it down a throat an infinite
spatial distance away. Hence the appearance of a new asymptotic region
is not avoided. Looking down the throat using the coordinate $\zeta$
of eq.~\reef{funnelz}, and inside the deSitter
horizon, the geometry is well approximated by
\beq\label{dSthroat}
ds^2\simeq\frac{1}{A^2(1-y)^2}\left[ F(y)d\hat{t}^2
-4\frac{dy^2}{F(y)}+2\left(d\zeta^2+e^{-2\zeta}d\phi^2\right)\right]\,,
\eeq
which is qualitatively similar to the case without a cosmological
constant, eq.~\reef{throat}.

As before, we are also interested in investigating the asymptotic future
evolution of the geometry. In this case we have to go beyond the future
deSitter horizon, to the region where $y_0<y<x$. There, the spacelike
future infinity characteristic of deSitter appears by taking $y\to x\in
[-1,1]$, where $x$ is now one of the coordinates parameterizing future
infinity. Close to the pole at $x=-1$ (\ie away from the throat) the
future asymptotic geometry approaches deSitter space. To see this,
consider $1+x\ll x-y\simeq 1+y$, and introduce new coordinates
\beq
1+x=\frac{A^2 r^2}{4}\,,\qquad
\frac{2H}{A^2}\hat{t}=w\,,\qquad 1+y=e^{-H\tau}\,,
\eeq
to recover exponentially expanding deSitter space
\beq
ds^2\to -d\tau^2+e^{2H\tau}\left(dw^2+dr^2+r^2d\phi^2\right)\,.
\eeq

Near the opposite pole at $x=+1$, we are in the throat region. The
future asymptotic metric
is also locally equivalent to deSitter space, which can be made manifest
by introducing, besides the coordinate $\zeta$ in \reef{funnelz}, new
coordinates (compare to \reef{tomilne})
\beq
w=\frac{\sqrt{2}}{A}\hat{t}\,,\qquad \sinh H\tau=\frac{\sqrt{2} H}{A(1-y)}
\eeq
in terms of which
\beq\label{aftermathds}
ds^2\to
-d\tau^2+\frac{1}{H^2}\sinh^2 (H\tau)\left(d\zeta^2+e^{-2\zeta}d\phi^2\right)
+\cosh^2 (H\tau) dw^2\,.
\eeq
If the coordinate $\phi$ were non-compact and ranged over the entire
real line, this would be equivalent to the portion of deSitter space
that lies beyond the cosmological horizon (an ``upper wedge'' akin to
the Milne region of Minkowski space). However, like in the previous
section, $\phi$ must be periodic by continuity of the full solution
across the horizon. Hence, the asymptotic geometry in the region close
to the axis that runs between the ghosts is deSitter space with a
periodic spatial circle. Of course, as $\tau\to\infty$ the exponential
expansion inflates the size of this circle, at any finite
value of $\zeta$, so eventually it becomes effectively non-compact.

The Euclidean instanton is obtained by Wick-rotating $\hat{t}\to i t_E$.
Regularity (up to the naked ghost singularity, which we assume
is smoothed out by ghost matter) requires the identification
\beq
t_E\sim t_E +\frac{8\pi}{F'(y_0)}\,,
\eeq
associated to the finite temperature of the deSitter horizon.

In order to compute the Euclidean action we use again \reef{actionarea},
which allows us to avoid dealing with the specific action for ghost
matter, as long as it satisfies the Einstein equations. There is no
contribution from the boundary of the infinite throat since the
Gibbons-Hawking term vanishes there. The acceleration horizons now are
actually cosmological horizons, which are finite without the need
for regularization. The reference background is deSitter space with the
same
value for $H$, and a simple calculation yields
\beq\label{donkeydsact}
I_E=\frac{\pi}{2G H^2}(1+y_0)\,.
\eeq
Since $1+y_0<0$, this action, like in the asymptotically flat case considered earlier, is
always negative, but the decay rate is exponentially suppressed as
$\Gamma\sim e^{-|I_E|}$.

To illustrate this result, we consider a small cosmological constant and
expand in powers of $H^2G^2m_-^2$ (\ie the square of the ratio between the
gravitational size of the lump and the Hubble radius). We find
\beq\label{donkeydsact2}
I_E\simeq -2\pi G m_-^2\left(1-4H^2G^2m_-^2+\dots\right)\,.
\eeq
In the limit $H\to0$ we correctly recover our previous result
\reef{donkeyact}.

\section{Discussion}
\label{sec:disc}

We have studied theories with ghosts (``the Others") that
interact very weakly, through gravity, with the ordinary world. Empty
space can decay into ghosts and ordinary particles, and we have focussed
on the possibility that non-relativistic lumps of
ghost matter be produced by tunneling processes, involving only
curvature scales smaller than $\mu$.

In ordinary field theory, nucleation rates are usually estimated by
using instanton methods, as $\Gamma\sim e^{-I_E}$, where $I_E$ is the
corresponding action. We have computed $I_E$ for a
variety of such possible decays. In some cases (eqs.~\reef{diwact},
\reef{donkeyact}, \reef{donkeydsact}), $I_E$ has turned out to be
negative. Naively, this would appear to lead to an exponentially
enhanced catastrophic decay rate. However, the Euclidean path integral
(even in the absence of gravity) in theories with ghosts is ill-defined,
and the standard Euclidean rules do
not apply. In section \ref{sec:ghostwalls} we have discussed nucleation
processes using the canonical WKB approach, without
reference to Euclidean methods. We have argued that for systems of a
single degree of freedom,
and for tightly bound systems of ghost and ordinary matter, the
nucleation rates should be estimated as
\beq
\Gamma \sim e^{-|I_E|}. \label{tura2}
\eeq
This formula may not be valid if, besides the gravitational interaction
between ghosts and matter, there are other dominant forces contributing
to the tunneling (such as the expansion of the universe). Here, we are
mainly interested in the case when ghosts and matter nucleate from
nearly flat space, and the tunneling is possible due to the
gravitational energy-momentum transfer between matter and ghosts. In
this situation, we expect (\ref{tura2}) to be valid. In more general
cases, we expect that Eq.~(\ref{tura2}) represents only a conservative
upper bound to the nucleation rate. Investigation of this more general
case is left for further research.

A first class of decays involves the nucleation in flat space of
lumps of ordinary matter together with their ghost counterparts.
For the spontaneous nucleation of diwalls (an
asymptotically flat configuration of concentrical walls, with an
ordinary wall chasing after a ghost wall), the corresponding Euclidean
action is given by Eq.~(\ref{diwact}). If the tensions of the two walls
are very different in magnitude, the action is of the form $|I_E| \sim
M_P^6/\sigma_1^2$, where $\sigma_1$ is the tension of the ordinary wall.
If both tensions are very similar, then the action is parametrically
given by
\beq\label{diwrate}
|I_E| \sim {M_P^6 \over \sigma^2}{\delta\sigma\over\sigma}\,.
\eeq
The curvature radius of the walls is given by $R\sim 1/ G\sigma$, so the
above formula takes the form $|I_E| \sim M_P^2 R \delta R$. The
effective theory (\ref{esym}) is supposed valid only at distance scales
larger than the cutoff $\mu^{-1}$, so within the regime of validity of
this effective theory,
\beq\label{extrasupp}
|I_E|\gsim \left(\frac{M_P}{\mu}\right)^2\simeq 10^{60}\,
\eeq
so the process of diwall nucleation is
extraordinarily suppressed, $\Gamma \sim e^{-10^{60}}$.

Such suppression does not appear to be at work in the nucleation
of a pair of Bondi dipoles (the non-perturbative analogue of
\reef{decay1}), for which we have computed the action to be
\beq\label{diprate} I_E\simeq 2\pi m_+ d\,, \eeq where $m_+$ and
$d$ are the mass and size of the dipole. There is no obvious
Planck-scale suppression here. However, since we require $d\gtrsim
\mu^{-1}$, the process will be suppressed for all lumps whose mass
$m$ is much larger than $\mu$. In the context of the Standard
Model with parity symmetry, the only elementary particles which
might have a mass comparable to the gravitational cut-off scale
are neutrinos. Could we then have non-perturbative nucleation of
neutrinos and ghost-neutrinos? The existence of the corresponding
instanton requires that the ordinary particle have larger mass
than its ghost counterpart, eq.~\reef{massdiff}. Now, the symmetry
between the normal and ghost sectors is broken only by their
different coupling to gravity. So any differences between the
masses of a particle and its ghost must be due to self-energy
corrections from the coupling to gravity. Since normal particles
are gravitationally attractive, whereas ghosts are repulsive, the
gravitationally-induced corrections to the mass will be \beq
(\Delta m_+^2)_\mathrm{grav}<0\,,\qquad (\Delta
m_-^2)_\mathrm{grav}>0\,. \eeq These work in a direction opposite
to \reef{massdiff}, and therefore the dangerous decay appears to
be not simply suppressed but actually forbidden!

Lastly, we have analyzed the non-perturbative analogue of \reef{decay2}.
This involves a dramatic topology change mediated by an
instanton with negative Euclidean action
\beq\label{ghpair}
I_E\simeq -2\pi G m_-^2\,.
\eeq
As we have seen, the negative sign does not spell doom since we should
use \reef{tura2} for the decay rate.
If we demand that the size of the nucleated configuration
$\sim A^{- 1}\sim Gm_-$ be larger than the cutoff scale $\mu^{-1}$, then this
process is as strongly suppressed as \reef{extrasupp}, and therefore
imposes no phenomenological constraints.

The decays of flat space involving ghost matter that we have studied do
not, by any means, exhaust all possibilities. We have considered ghost
domain walls and ghost particles, but ghost cosmic strings can also
unstabilize the vacuum. A ghost cosmic string coupled to gravity creates
a conical excess. If the string is open with ghost monopoles at its
endpoints (\ie the string is topologically unstable) then the string
will pull the monopoles together: such strings can not pop out of the
vacuum spontaneously. However, string vortices, even topologically
stable ones, can end on black holes \cite{agk}. If a ghost string has
black holes (which necessarily have positive mass) at its endpoints,
then the string will push the black holes apart. One can then envisage a
process in which one such ghost string with two black holes attached is
nucleated out of the flat vacuum. Following \cite{strbh}, the relevant
instanton can be easily constructed, and the action is
\beqa\label{ghstring}
I_E\simeq \frac{\pi m^2}{T}-\pi Gm^2
=\frac{\pi m^2}{T}\left(1-\frac{T}{M_P^2}\right)\,,
\eeqa
where $m$ is the mass of the black holes and $T$ the absolute value of
the ghost string tension. The first contribution in \reef{ghstring} is
the nucleation rate of the string itself, and the second is an
enhancement of the decay due to black hole entropy\footnote{In order to
construct the instanton the black holes must charged and extremal (or
almost extremal) \cite{strbh}, so their entropy is $S_{bh}\simeq \pi
Gm^2$.}. The latter becomes subleading for string tensions $T\ll M_P^2$,
so the factor in brackets in \reef{ghstring} can be taken to be of order
one.
In order that the size of the nucleated configuration be
larger than the gravitational cutoff we must require $m/T\gsim\mu^{-1}$.
Since the black holes must also be larger than Planck size, $m\gsim
M_P$, we find that
\beq
I_E \gsim \frac{M_P}{\mu}\sim 10^{30}\,.
\eeq
Again, the nucleation rate is strongly suppressed.

The results presented in this paper are quite encouraging for the
energy-parity-symmetric theory in (\ref{esym}). In all the cases where
we have found the possibility to conjure ghosts, we have seen they can
be exorcized away for all of eternity (for practical purposes), in
the sense that $|I_E|$ is always extremely large for the
non-perturbative decay processes considered. Nevertheless, it
should be kept in mind that the Euclidean path integral is ill-defined
in theories with a ghost sector, and our justification of
Eq.~(\ref{tura2}) for the nucleation rates is rather heuristic. In this
sense, our analysis should be regarded as preliminary. A more rigorous
derivation of nucleation rates in theories with ghosts deserves further
investigation. One might also worry that the suppression of the decay
seems to be of a different sort in each of the cases we have studied:
while \reef{diwrate}, \reef{ghpair}, \reef{ghstring} are suppressed by
the weakness of gravity, the Bondi pair prodution \reef{diprate} is
instead eliminated by dynamical considerations. There does not seem to
be an underlying generic argument that guarantees that ghost decays are
always phenomenologically harmless. Here we have considered what appear
to be the simplest and most natural decay channels, but we cannot rule
out the existence of a different and dangerous non-perturbative
instability.

\section*{Acknowledgements}
This work is partially supported by grants
CICyT FPA2004-04582-C02-02, European Comission FP6 program
MRTN-CT-2004-005104, and DURSI 2005SGR 00082.

\section*{Appendix: Tunneling paths}

In Section 2.4 we considered the nucleation of walls and diwalls
using the canonical WKB approach. Although the explicit form of
the tunneling path was not needed in order to integrate the
Hamilton-Jacobi equation, it remains to be shown that such path
can be constructed. Also , in order to remove a certain sign
ambiguity [see the discussion around Eq.~(\ref{signs})] we assumed
that there is a path where the momentum associated with the
position of the wall vanishes, $p=0$, or where $V_+=V_-$. Here, we
discuss the construction of such paths. Again, we follow the
approach of Ref.~\cite{fmp} (with minor modifications to
accommodate the case of compact geometries).

Eqs. (\ref{pandq}) with $p=0$ can easily be solved for $\hat
R'_\pm \equiv R'(r=q\pm \epsilon)$: \beq {\hat R'_\pm\over L} =
{1\over 2}\left({(V_--V_+) \over 4\pi G \sigma \hat R} \mp {4 \pi
G \sigma \hat R}\right). \label{pau}\eeq Likewise, from
(\ref{pil}), we can find the momentum $\hat \Pi_L \equiv
\Pi_L(r=q)$, \beq \hat \Pi_L^2 = {\hat R^2 \over G^2}
\left({(V_+-V_-)^2 \over 64\pi^2 G^2 \sigma^2 \hat R^2} - {V_++V_-
\over 2}+ 4 \pi^2 G^2 \sigma^2 \hat R^2 - 1 \right).\eeq

For instance, in the case when the cosmological constants are the
same on both sides of the wall, we can take the turning point
before tunneling to be regular inside the wall, so that $M_-=0$,
but allow for an infinitessimal mass of the initial bubble $M_+
\neq 0$. In this case $V_-=1-H^2 \hat R^2$ and $V_+=1-H^2\hat R^2
- 2 GM_+ /\hat R$. The momentum is given by
$$
\hat \Pi_L^2 = {\hat R^2\over G^2}\left\{H^2\hat R^2+\left(2\pi
G\sigma \hat R + {M_+ \over 4\pi\sigma \hat R^2}\right)^2 -1
\right\}.
$$
The zeroes of the quantity in curly brackets define the turning
points. In the limit when $M_+ \to 0$, one of the turning points
tends to zero (vanishing initial size)
$$
\hat R_b \approx (M_+/4\pi\sigma)^{1/2} \to 0,
$$
and the other is given by the size of the wall at the moment of
nucleation
$$
\hat R_a = R_w = [H^2 + (2\pi G\sigma^2)]^{-1/2}.$$ In this limit,
the expression simplifies to $ \hat \Pi_L^2 = G^{-2}\hat R^2
[1-(H^2+(2\pi G\sigma )^2) \hat R^2]. $

Let us first consider the case of negative tension walls
nucleating in deSitter. For given radius of the wall $\hat R$, we
can find a three geometry by solving the following differential
equations:
\begin{eqnarray}
\left({R'\over L}\right)^2 &=& 1+ V(R) + {G^2\ \hat\Pi_L^2\over
\hat R^2} \left({R\over \hat R}\right)^{2},   \quad\quad (r<q), \label{rp1}\\
\left({R'\over L}\right)^2 &=& (1+V(R))\left\{ 1 +{G^2\
\hat\Pi_L^2\over \hat R^2 (1+ V_+(\hat R))}\right\}.\quad\quad
(r>q). \label{rp2}
\end{eqnarray}
These 3-geometries will automatically satisfy the appropriate
junction conditions at $R=\hat R$. They will also correspond to
the turning point geometries when $\hat\Pi_L=0$, that is, when
$\hat R= \hat R_a$ or $\hat R = \hat R_b$. For intermediate values
of $\hat R$, they interpolate between turning points.
The sequence of 3-geometries is schematically represented in Fig. \ref{fig:ghostpath}.
It is
important that for $R\to 0$ we have $R'/L \to 1$, so that the
geometry is locally flat at the origin, inside the wall. On the
other hand, right outside the wall, Eq.~(\ref{pau}) shows that the
radius is growing $\hat R'_+ >0$, hence it is natural to choose a
boundary condition such that as we approach the horizon $1+V=0$,
we have $R'=0_+$. This geometry can either be continued
analytically past the horizon, or simply matched smoothly with the
original turning point solution beyond the horizon (leaving the
part beyond the horizon unchanged throughout the whole
semiclassical path.) For $r<q$, we have $V=-H^2R^2$, and Eq.
(\ref{rp1}) leads to
\beq
R(y) = \alpha^{-1} \cos \alpha
y,\quad\quad (r<q),
\eeq
where $y$ stands for proper distance, $dy
= L dr$, and
\beq
\alpha^2 \equiv H^2 - {G^2\ \hat\Pi_L^2\over
\hat R^4} > H^2.\label{mole}
\eeq
In the last inequality, we have
used that $\hat\Pi_L^2 < 0$ under the barrier. Thus, the geometry
inside of the wall is just that of a cap of a three-sphere of
radius $\alpha^{-1}$. At the first turning point, when $\hat R=
\hat R_b \approx (M_+/4\pi\sigma)^{1/2}$, $\hat \Pi_L$ vanishes
and the curvature radius is ``large", $\alpha^{-1} = H^{-1}$.
However, as $\hat R$ grows to be a few times $\hat R_a$, we find
that the second term in (\ref{mole}) quickly dominates, and the
curvature radius of the three sphere becomes much smaller,
$\alpha^{-1} \sim (M_+/\sigma)^{1/2}$ (vanishingly small as $M_+$
tends to zero). The curvature radius of the geometry inside the
bubble grows back to its original size $\alpha^{-1} = H^{-1}$ as
the second turning point is approached. The solution outside the
bubble is obtained by integrating Eq.~(\ref{rp2}). Note that the
term in curly brackets in the right hand side is just a constant,
and so the solution is basically the warp factor of an
``Anti"-Schwarzschild-deSitter geometry (with negative mass
parameter $M_+ < 0$), with ``Hubble" parameter $H$, but with an
unconventional parametrization. For instance, in the limit $M_+\to
0$ we have
$$
R(y)\approx H^{-1}\cos\beta H y, \quad\quad (r>q),
$$
where $\beta < 1$ is the square root of the term in curly brackets
in Eq.~(\ref{rp2}). The 3-geometry always contains a maximal
circle of the size of the horizon of the (fictitious)
``Anti"-Schwarzschild-deSitter geometry. In particular, the
3-volume is large for any value of $\hat R$. In the limit $M_+\to
0$ (and ignoring the fact that a small part of the volume is cut
out by the bubble) the 3-volume scales like $V_3 \sim 1/\beta
H^3$. Hence, we have found a sequence of three geometries
interpolating between both turning points, where the 3-volume is
always of the order $H^{-3}$ or larger. This is in contrast with
the picture where the three geometry first disappears and then
reappears with a large bubble in it.

For walls of positive tension, we can construct interpolating
geometries along the same lines. However, for  Eq.~(\ref{pau}) now
shows that in the limit $M_+\to 0$ and finite $\hat R$, the radius
$R(r)$ is always decreasing right outside the bubble $\hat R'_+ <
0$. Hence, it is convenient to choose the boundary condition for
$r>q$ in a form appropriate to decreasing radii. Thus, instead of
(\ref{rp2}) we can use (\ref{rp1}) also outside the bubble.
The sequence of 3-geometries is schematically represented in Fig. \ref{fig:wallpath}.
Initially, in the limit of small mass $GM_+ \ll H^{-1}$, the
geometry is basically that of a 3-sphere of radius $H^{-1}$ with a
small spherical wall of radius $\hat R=\hat R_b \sim
(M_+/\sigma)^{1/2}.$ As we increase $\hat R$ to a few times $\hat
R_b$, the curvature radius of the geometry on both sizes of the
wall becomes comparable to the radius of the wall. As $\hat R$ is
further increased, the geometry on both sides becomes more and
more symmetric, corresponding to two spherical caps of curvature
radius $\alpha^{-1}$. As we approach the second turning point,
$\hat R = \hat R_b = R_w$, this curvature radius becomes $H^{-1}$.
In this case, the path we are considering first shrinks the
spatial geometry to a very small size $\sim (M_+/\sigma)^{1/2}$,
and then the new geometry grows from this seed. Thus, in the
present case, if we take the limit when $M_+ \to 0$, the initial
geometry disappears and the new one reappears from ``nothing".

Finally, we may consider the paths for nucleation of diwalls. In
this case, the construction of a path with $p=0$ seems rather
tricky, because there are two independend wall radii. Instead,
since the instanton in this case has the trivial topology of
$R_4$, we can construct a semiclassical path by simply slicing the
instanton on surfaces of constant coordinate $T\equiv y \cos\chi$,
where $\chi$ is the polar coordinate on the three-spheres. For
$T\to -\infty$ the geometry is that of flat three-dimensional
Euclidean space, while for $T=0$ we have the turning point with
two nested domain walls. The potential vanishes everywhere, $V=0$,
and hence the equality $V_+=V_-$ is trivially satisfied accross
each wall. This justifies the use of Eq.~(\ref{signs}) in this
case too.


\begin{thebibliography}{99}



\bibitem{kasu}
D.~E.~Kaplan and R.~Sundrum,
  ``A symmetry for the cosmological constant,''
  arXiv:hep-th/0505265.
  %%CITATION = HEP-TH 0505265;%%

\bibitem{li}
A.~D.~Linde,
  ``The Inflationary Universe,''
  Rept.\ Prog.\ Phys.\  {\bf 47} (1984) 925.
  %%CITATION = RPPHA,47,925;%%

\bibitem{de}
R.~R.~Caldwell,
  ``A Phantom Menace?,''
  Phys.\ Lett.\ B {\bf 545} (2002) 23
  [arXiv:astro-ph/9908168].
  %%CITATION = ASTRO-PH 9908168;%%

\bibitem{ca}
S.~M.~Carroll, M.~Hoffman and M.~Trodden,
  ``Can the dark energy equation-of-state parameter w be less than -1?,''
  Phys.\ Rev.\ D {\bf 68} (2003) 023509
  [arXiv:astro-ph/0301273].
  %%CITATION = ASTRO-PH 0301273;%%

\bibitem{cline}
  J.~M.~Cline, S.~Jeon and G.~D.~Moore,
  ``The phantom menaced: Constraints on low-energy effective ghosts,''
  Phys.\ Rev.\ D {\bf 70} (2004) 043543
  [arXiv:hep-ph/0311312].
  %%CITATION = HEP-PH 0311312;%%


\bibitem{an} %\cite{Anisimov:2005ne}
  A.~Anisimov, E.~Babichev and A.~Vikman,
  ``B-inflation,''
  JCAP {\bf 0506} (2005) 006
  [arXiv:astro-ph/0504560].
  %%CITATION = ASTRO-PH 0504560;%%

\bibitem{bondi}
H.~Bondi,
``Negative Mass in General Relativity,"
Rev.\ Mod.\ Phys.\ {\bf 29}, 423 (1957).

\bibitem{iskh}
W.~Israel and K.~A.~Khan, ``Collinear particles and Bondi dipoles
in General Relativity," Nuovo.\ Cim.\ {\bf 33}, 331 (1964).

\bibitem{bonnor}
W.~B.~Bonnor and N.~S.~Swaminarayan, ``An exact stationary
solution of Einstein's equations,'' Z.\ Phys.\ {\bf 186}, 222
(1965).


\bibitem{gibbons}
G.~W.~Gibbons,
  ``The motion of black holes,''
  Commun.\ Math.\ Phys.\  {\bf 35} (1974) 13.
  %%CITATION = CMPHA,35,13;%%

\bibitem{av} A.~Vilenkin,
  ``Gravitational field of vacuum domain walls,''
  Phys.\ Lett.\ B {\bf 133} (1983) 177.
  %%CITATION = PHLTA,B133,177;%%

\bibitem{di} W.~Israel,
  ``Singular hypersurfaces and thin shells in General Relativity,''
  Nuovo Cim.\ B {\bf 44S10} (1966) 1
  [Erratum-ibid.\ B {\bf 48} (1967\ NUCIA,B44,1.1966) 463].
  %%CITATION = NUCIA,B44S10,1;%%

\bibitem{bgv} R.~Basu, A.~H.~Guth and A.~Vilenkin,
  ``Quantum creation of topological defects during inflation,''
  Phys.\ Rev.\ D {\bf 44}, 340 (1991).
  %%CITATION = PHRVA,D44,340;%%

\bibitem{fmp} W.~Fischler, D.~Morgan and J.~Polchinski,
  ``Quantization Of False Vacuum Bubbles: A Hamiltonian Treatment Of
  Gravitational Tunneling,''
  Phys.\ Rev.\ D {\bf 42} (1990) 4042.
  %%CITATION = PHRVA,D42,4042;%%



\bibitem{garyrob}
G.~T.~Horowitz and R.~C.~Myers,
  ``The value of singularities,''
  Gen.\ Rel.\ Grav.\  {\bf 27}, 915 (1995)
  [arXiv:gr-qc/9503062].
  %%CITATION = GR-QC 9503062;%%

\bibitem{synge}
J.~L.~Synge, ``Relativity: The General Theory" (North Holland,
Amsterdam, 1960).


\bibitem{fay}
H.~F.~Dowker and S.~N.~Thambyahpillai,
  ``Many accelerating black holes,''
  Class.\ Quant.\ Grav.\  {\bf 20} (2003) 127
  [arXiv:gr-qc/0105044].
  %%CITATION = GR-QC 0105044;%%

\bibitem{er}
R.~Emparan and H.~S.~Reall,
  ``Generalized Weyl solutions,''
  Phys.\ Rev.\ D {\bf 65} (2002) 084025
  [arXiv:hep-th/0110258].
  %%CITATION = HEP-TH 0110258;%%


\bibitem{kw}
W.~Kinnersley and M.~Walker,
  ``Uniformly Accelerating Charged Mass In General Relativity,''
  Phys.\ Rev.\ D {\bf 2}, 1359 (1970).
  %%CITATION = PHRVA,D2,1359;%%

\bibitem{bonnorC}
W.~B.~Bonnor,
``The sources of the vacuumC-metric,"
Gen.\ Rel.\ Grav.\ {\bf 15}, 535 (1983).

\bibitem{gamow}
G.~Gamow, ``Biography of Physics" (Harper \& Row, New York, 1961).

\bibitem{zeromass}
R.~Emparan,
  ``Massless black hole pairs in string theory,''
  Phys.\ Lett.\ B {\bf 387} (1996) 721
  [arXiv:hep-th/9607102];
  %%CITATION = HEP-TH 9607102;%%
``Composite black holes in external fields,''
  Nucl.\ Phys.\ B {\bf 490} (1997) 365
  [arXiv:hep-th/9610170].
  %%CITATION = HEP-TH 9610170;%%



\bibitem{HaHo}
S.~W.~Hawking and G.~T.~Horowitz,
  ``The Gravitational Hamiltonian, action, entropy and surface terms,''
  Class.\ Quant.\ Grav.\  {\bf 13}, 1487 (1996)
  [arXiv:gr-qc/9501014].
  %%CITATION = GR-QC 9501014;%%

\bibitem{ht}
K.~Hong and E.~Teo,
  ``A new form of the C-metric,''
  Class.\ Quant.\ Grav.\  {\bf 20}, 3269 (2003)
  [arXiv:gr-qc/0305089].
  %%CITATION = GR-QC 0305089;%%

\bibitem{emgu}
R.~Emparan and M.~Gutperle,
  ``From p-branes to fluxbranes and back,''
  JHEP {\bf 0112}, 023 (2001)
  [arXiv:hep-th/0111177].
  %%CITATION = HEP-TH 0111177;%%

\bibitem{pd}
J.~F.~Plebanski and M.~Demianski,
  ``Rotating, Charged, And Uniformly Accelerating Mass In General Relativity,''
  Annals Phys.\  {\bf 98}, 98 (1976).
  %%CITATION = APNYA,98,98;%%

\bibitem{agk}
A.~Achucarro, R.~Gregory and K.~Kuijken,
  ``Abelian Higgs hair for black holes,''
  Phys.\ Rev.\ D {\bf 52} (1995) 5729
  [arXiv:gr-qc/9505039].
  %%CITATION = GR-QC 9505039;%%

\bibitem{strbh}
S.~W.~Hawking and S.~F.~Ross,
  ``Pair production of black holes on cosmic strings,''
  Phys.\ Rev.\ Lett.\  {\bf 75}, 3382 (1995)
  [arXiv:gr-qc/9506020].
  %%CITATION = GR-QC 9506020;%%
\\
R.~Emparan,
  ``Pair creation of black holes joined by cosmic strings,''
  Phys.\ Rev.\ Lett.\  {\bf 75}, 3386 (1995)
  [arXiv:gr-qc/9506025].
  %%CITATION = GR-QC 9506025;%%
\\
D.~M.~Eardley, G.~T.~Horowitz, D.~A.~Kastor and J.~H.~Traschen,
  ``Breaking cosmic strings without monopoles,''
  Phys.\ Rev.\ Lett.\  {\bf 75}, 3390 (1995)
  [arXiv:gr-qc/9506041].
  %%CITATION = GR-QC 9506041;%%


\end{thebibliography}
\end{document}